\PassOptionsToPackage{pagebackref,breaklinks,colorlinks}{hyperref}
\PassOptionsToPackage{table}{xcolor}

\documentclass{article}
\usepackage[main, final]{neurips_2026}
\usepackage[utf8]{inputenc}
\usepackage[T1]{fontenc}
\usepackage{booktabs}
\usepackage{amsfonts}
\usepackage{comment}
\usepackage{nicefrac}
\usepackage{microtype}
\microtypecontext{spacing=nonfrench}
\usepackage{pifont}
\usepackage{graphicx}
\usepackage{amsmath}
\usepackage{caption}
\usepackage{subcaption}
\usepackage{multirow}
\usepackage{makecell}
\usepackage{algorithm2e}
\usepackage{enumitem}
\usepackage{wrapfig}
\usepackage{tikz}
\usepackage[most]{tcolorbox}
\usepackage{url}
\usepackage{amssymb}
\usepackage{type1cm}

\definecolor{coralred}{RGB}{220,80,60}
\definecolor{lightgreen}{RGB}{230,242,222}
\definecolor{lightgreen2}{RGB}{241,248,235}
\definecolor{lightgreen3}{RGB}{236,245,228}
\definecolor{lightyellow}{RGB}{255,255,210}
\definecolor{lightorange}{RGB}{255,235,210}
\definecolor{default}{RGB}{255,255,255}
\definecolor{pseudoblue}{RGB}{55,113,175}
\definecolor{mgreen}{RGB}{6,128,67}
\definecolor{mgray}{RGB}{240,240,240}
\definecolor{mgray_dark}{RGB}{128,128,128}
\definecolor{mygreen}{RGB}{233,247,234}
\definecolor{mygray}{gray}{0.97}
\definecolor{purple}{RGB}{128,128,247}
\definecolor{mypink}{RGB}{252,237,236}
\definecolor{pink2}{RGB}{234,223,222}
\definecolor{pink3}{RGB}{237,227,228}
\definecolor{blue1}{RGB}{229,243,247}

\NewDocumentCommand{\circnumgreen}{O{teal!18} m}{%
  \tikz[baseline=(char.base)]{
    \node[shape=circle, fill=#1, inner sep=1pt, font=\small\bfseries] (char)
      {\textcolor{teal!70!black}{#2}};
  }%
}

\NewDocumentCommand{\circnumpink}{O{pink!30!white} m}{%
  \tikz[baseline=(char.base)]{
    \node[shape=circle, fill=#1, inner sep=1pt, font=\small\bfseries] (char)
      {\textcolor{pink!70!red}{#2}};
  }%
}

\newcounter{findingcnt}

\newtcolorbox{redquestion}[1][]{
  enhanced,
  boxrule=0pt,
  leftrule=4pt,
  frame hidden,
  colback=mypink,
  colframe=coralred,
  sharp corners,
  left=3pt,
  right=3pt,
  top=-1.5pt,
  bottom=-1.5pt,
  boxsep=1pt,
  before skip=1pt,
  after skip=1pt,
  parbox=false,
  before={\par\vspace{-2pt}\noindent},
  after={\par\vspace{-4pt}},
  before upper={
    \stepcounter{findingcnt}
    \textbf{\textcolor{coralred}{Finding \#\thefindingcnt}}~%
  },
  #1
}

\begin{document}

\date{}

\title{\Large \bf Rethinking Latency Denial-of-Service: Attack the LLM Serving Framework, Not the Model}

\author{
Tianyi Wang, 
Huawei Fan, 
Yuanchao Shu, 
Peng Cheng, 
Cong Wang\thanks{Corresponding Author}\\
Zhejiang University, Hangzhou, China\\
{\tt\small \{wty1998, hw.phil.fan, ycshu, lunarheart, cwang85\}@zju.edu.cn}
}

\maketitle

\begin{abstract}
LLM inference is inherently expensive, even a modest slowdown can translate into substantial operating costs and severe availability risks. Recently, a growing body of research known as latency attacks focuses on crafting inputs to trigger worst-case output lengths. However, we report a contrary finding that these algorithmic-level latency attacks are largely ineffective against modern LLM serving systems. We reveal that system-level optimization such as continuous batching provides a logical isolation to mitigate contagious latency impact on co-located users. Thus, in this paper, we shift our focus from the algorithm to the system layer, and introduce a new Fill and Squeeze attack strategy targeting the state transition of the scheduler. ``Fill'' first exhausts the global KV cache to induce Head-of-Line blocking, while ``Squeeze'' forces the system into repetitive preemption. By manipulating output lengths using different attack prompts, and leveraging side-channel probing of memory status, we demonstrate that the attack can succeed in a practical black-box setting with much less cost. Extensive evaluations on vLLM indicate up to $75-742\times$ TTFT degradation relative to benign baselines and $1.5-4\times$ average slowdown on Time Per Output Token compared to existing attacks with 30-40\% lower attack cost. 
Code: \url{https://github.com/Phil-Fan/FS-attack}
\end{abstract}

\vspace{-0.1in}
\section{Introduction}

LLM API spending has surged from \$3.5B to \$8.4B in the last six months~\cite{menlo2025},
driven by billion-scale daily requests~\cite{techcrunch_2025_chatgpt_2_5b_prompts}. Meeting strict latency requirements of conversational agents~\cite{schulman2022chatgpt,gpt}, code generation~\cite{tong2024codejudge,chen2021evaluating}, and
document summarization~\cite{wang2023element,zhang2024benchmarking} while maximizing GPU
utilization is the core challenge of modern LLM serving. As AI companies are competing to accommodate larger context windows (e.g., 1M/65K input/output tokens for Gemini 3.0 Pro~\cite{google_vertexai_gemini30pro_docs}), a new class of Denial-of-Service threats is looming known as \emph{latency attacks}~\cite{engorgio, LoopLLM,SpongeExamples,verbose_imgs,nicgslowdown,OverThink,autodos,ExtendAttack} that exploits the autoregressive nature of LLMs to suppress the end-of-sequence~(\texttt{EOS}) token and trap models in indefinite babbling. Existing literature reports $7-10\times$ latency amplification~\cite{verbose_imgs} that could \emph{theoretically} threaten service availability.

However, counter-intuitively, we first find that the existing model-centric latency attacks are surprisingly \textit{ineffective} in production environments. Through extensive empirical and theoretical analysis, we find that simply prolonging generation length fails to inflict significant latency on co-located users, and often leaves Time-to-First-Token (TTFT) nearly unchanged even when malicious requests constitute a large fraction of the workload. To answer ``\emph{Why latency attacks fail to delay?}'', we delve into the natural resilience of \emph{Continuous Batching} (CB)~\cite{vllm,orca,sglang,li2023alpaserve}, a technique originally proposed to schedule requests at the iteration level so that short requests do not need to wait for the long requests to finish. This helps create a \textit{logical isolation} between the normal traffic and the attacker's long-running request. Even if the attacker could occupy the scheduling slot, it does not prevent shorter, legitimate requests from entering and exiting the system. As a result, this relegates most latency attack as largely self-inflicted rather than contagious on co-located users~\cite{yan2025bithydra}.

To make latency attacks effective again, we shift our focus to the system layer by introducing \textbf{Fill and Squeeze} (F\&S), a new attack strategy targeting the deterministic state transitions of the scheduler. It can be integrated with any attack methods as pluggable payloads. The core intuition is to manipulate memory resources to trigger pathological behaviors rather than simply slow down individual generations. The proposed attack operates via two vectors: 1) \textit{Fill}, which focuses on exhausting the VRAM rapidly. By injecting adversarial requests that generate long sequences and potentially leveraging ambient traffic peaks~\cite{vijaya2025aqua}, the attacker could quickly drive the global KV cache usage to a boundary point that triggers the scheduler's admission control, inducing memory-based Head-of-Line (HOL) blocking; 2) \textit{Squeeze}, which exploits the preemption logic. Once the system is brought to the saturation boundary, it forces the scheduler into an alternation between continuous preemption and recovery, that diverts precious GPU cycles into expensive memory swapping or recomputation operations~\cite{vllm_optimization_config_stable, xu2024pie}. 

Yet, implementing this strategy in a production environment faces two challenges. The first one is the visibility gap. Different from white-box KV cache estimation~\cite{helix}, the attacker acts as a standard tenant with no privileged access to system internals (current KV cache occupancy). Hence, it is difficult to determine the exact timing to launch an injection and whether the injection has brought the system to the boundary proximity of GPU VRAM. Second is the economic viability. A brute-force strategy of continuously sending maximum-length requests incurs high monetary costs and low stealthiness~\cite{attack_detect,phute2023llm}. In the real world, where attackers must pay per token, an attack is only worthwhile if the damage outweighs the cost. Hence, it is necessary to come up with a more cost-effective strategy that can take advantage of the deterministic states of the serving scheduler. 

To tackle these challenges, we first discover that the Inter-Token Latency (ITL) of standard requests exhibits a strong correlation with global GPU memory usage. We leverage this proxy to train a lightweight regressor~\cite{ke2017lightgbm} that allows the attacker to estimate KV usage without system privileges in black-box. Further, to minimize attack cost, we formulate the attack into a constrained optimization problem, decompose attack prompts into three different tiers measured by output lengths and design a dynamic injection strategy to switch between the Fill and Squeeze at different intensities. The results show $20-280\times$ average slowdown on Time to First Token (TTFT) and $1.5-4\times$ average slowdown on Time Per Output Token (TPOT) compared to the existing attacks.

\vspace{-0.1in}
\section{Background and Related Works} \label{sec:background}

\vspace{-0.1in}
\subsection{LLM Serving}

LLMs generate text autoregressively via prefill and decoding, in which decoding continues until an end-of-sequence [\texttt{EOS}] token is produced or a predefined maximum length is reached. Because each decode step must read the accumulated KV-cache, it is \emph{memory-bound}. Production serving systems like vLLM~\cite{vllm}, Orca~\cite{orca}, SGLang~\cite{sglang} maximize throughput via Continuous Batching (CB), which schedules requests at the iteration level~\cite{orca,fairness}. When any running request completes, a new waiting request is immediately admitted, preventing short requests from queuing behind long ones. When aggregated demand exceeds capacity, the scheduler either \emph{preempts} active requests (via swap to CPU DRAM or KV recomputation) or blocks new admissions. 

To handle real-world dynamic request lengths, Sarathi-Serve~\cite{agrawal2024taming} proposes chunked prefill to split long prompts into smaller batches, piggybacking them onto decoding steps to reduce pipeline bubbles. DynaServe~\cite{dynaserve} introduces a unified execution plan that dynamically partitions resources between prefill and decode to handle unbalanced workloads. Similarly, Aqua decouples memory and compute allocation via network-accelerated memory offloading~\cite{vijaya2025aqua}. To address user fairness, VTC employs fair queuing to ensure equal resource distribution~\cite{fairness}. Importantly, there is a distinction between handling natural long-context traffic and malicious latency attacks. In production environments, the arrival of long requests follows a stochastic distribution (e.g., Poisson process)~\cite{fairness,vllm,dynaserve,sun2024llumnix}. Modern schedulers are robust to handle occasional congestion by swapping them to DRAM or delaying their admission until resources free up. In contrast, our work approaches LLM serving from an adversarial perspective by transforming the VRAM capacity constraint into a failure mode that standard scheduling policies (like FCFS or LIFO preemption) cannot easily resolve. 

\vspace{-0.05in}
\subsection{Latency Attacks}
Latency attacks on AI systems evolve under the classic DoS principle to induce computational overload. The pioneering examples are the series works of Sponge Examples~\cite{SpongeExamples,nicgslowdown,shapira2023phantom}, which targets NLP/vision models to deliberately inflate inference time and energy usage. Latency attacks can be broadly categorized into: optimization-based, whose objective is to postpone the generation of \texttt{EOS} token for as long as possible~\cite{engorgio, LoopLLM,LingoLoop_Attack,ExcessiveReasoningAttack}; prompt engineering-based, which exploits the model's instruction-following capability to maximize the output length~\cite{OverThink, autodos, ExtendAttack}. It is achieved by constructing prompts that involve complex logical reasoning or force the model to explain multiple concepts. Although the existing attacks are theoretically sound, we demonstrate that most of these model-centric approaches struggle in production environments. In contrast, this work targets the scheduler by exploiting the resource management logic rather than optimizing the generation length.

\section{Why Latency Attacks Fail under LLM Serving}
\label{sec:motivation}

\subsection{Empirical Analysis}

\begin{wrapfigure}{r}{0.4\linewidth}
    \centering
    \setlength{\abovecaptionskip}{2pt}
    \includegraphics[width=\linewidth]{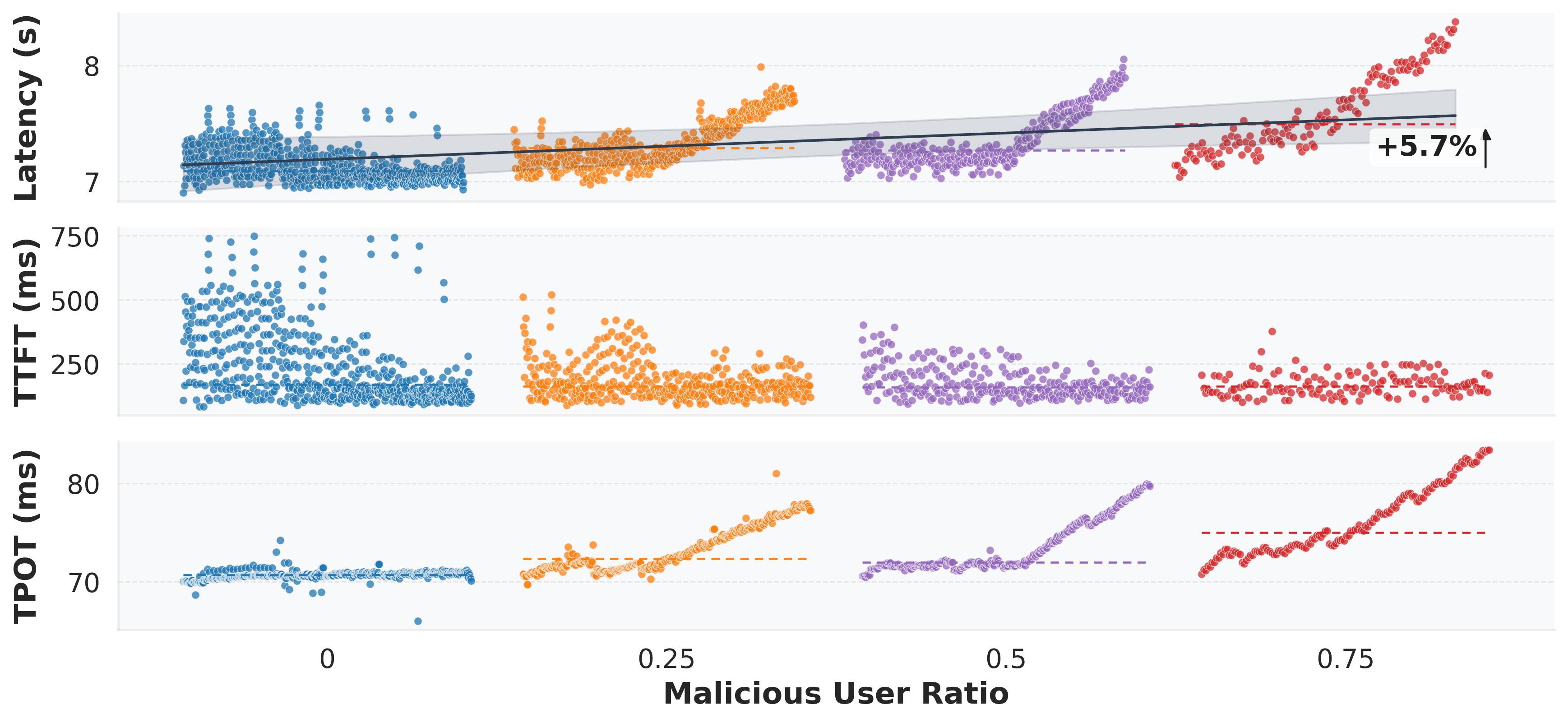}
    \caption{\small{Impact of latency attacks on normal users.}}
    \label{fig:degradation}
\end{wrapfigure}

We first perform empirical analysis by establishing a multi-tenant environment on the vLLM framework (NVIDIA L40S, 48GB VRAM running Qwen3-8B~\cite{qwen3}, \texttt{gpu\_memory\_utilization}=$0.90$) with batch size $16$ and varying proportions of adversarial requests. 
\begin{redquestion} \textbf{(CB Naturally Defends Against Latency Attacks).}  As shown in Fig. \ref{fig:degradation}, even with 50-75\% adversarial requests using LoopLLM (a strong attack)~\cite{LoopLLM}, the TTFT of normal users increases by \emph{only} $3-5.7$\% and TTFT remains nearly flat.  
\end{redquestion}
\setlength{\intextsep}{0pt}       
This finding represents a sharp contrast to the existing works claiming over orders of magnitude of latency~\cite{verbose_imgs}. We attribute such natural resilience to CB, which was originally designed to improve throughput. Unlike static batching, where batch latency is dictated by the longest sequence (HOL blocking), CB dynamically schedules requests at the iteration level. 
An attacker's request that delays emitting an \texttt{EOS} token simply occupies scheduling slot for a longer duration, but it does not logically block the scheduling of shorter, legitimate requests, which can enter and exit the batch independently. Thus, model-centric latency attacks become self-inflicted rather than contagious. 

\begin{wrapfigure}{r}{0.4\linewidth}
     \centering
     \includegraphics[width=\linewidth]{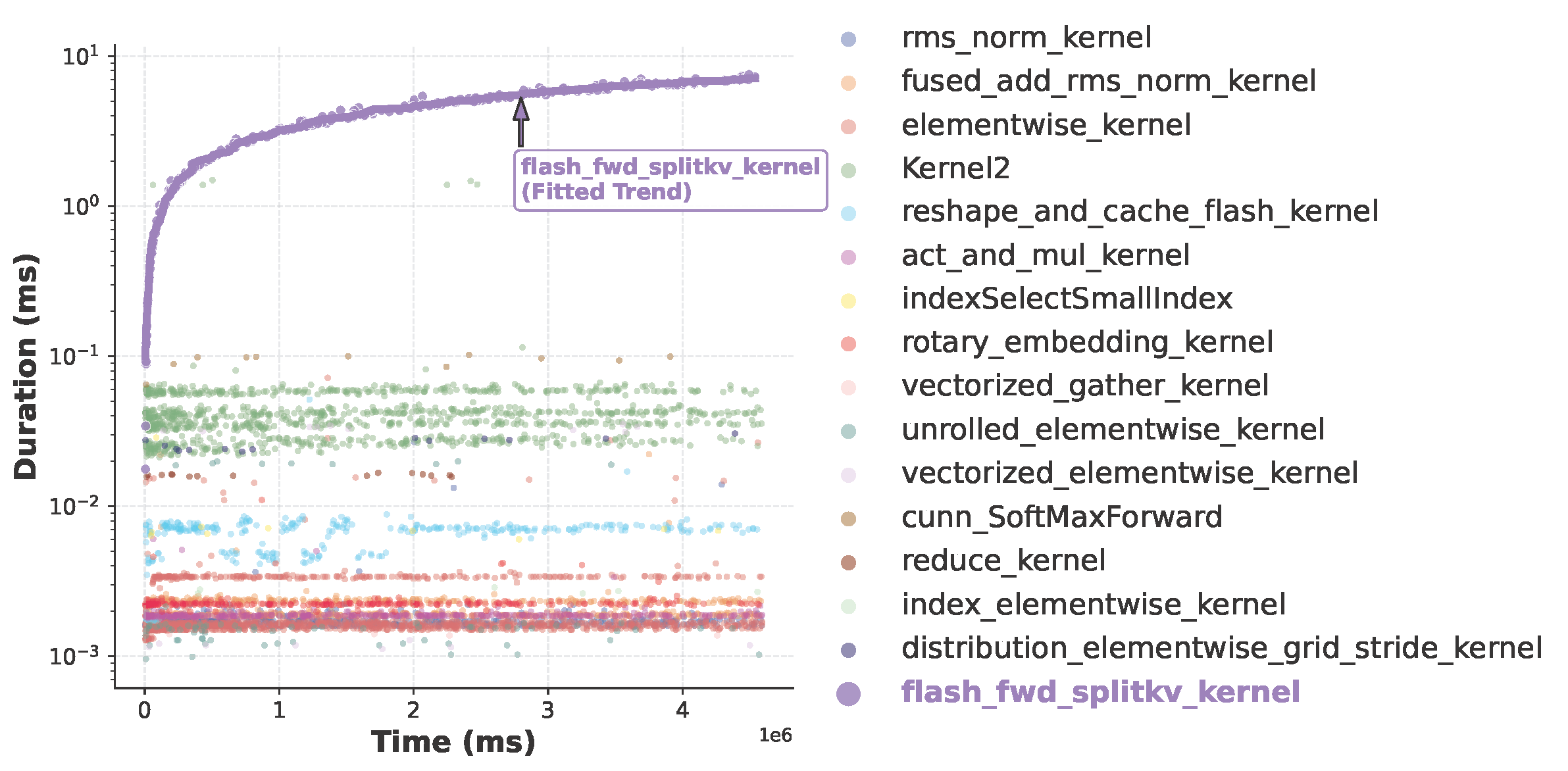}
     \caption{\small{CUDA kernel execution under the latency attack.}} 
     \label{fig:kernel_func}
 \end{wrapfigure}

\begin{redquestion} \textbf{(TPOT Leaks Physical Contention).} While TTFT is protected, there is a consistent TPOT increase of 5ms (roughly 7\%) under adversarial load in Fig. \ref{fig:degradation}. CUDA profiling (Figure \ref{fig:kernel_func}) reveals \texttt{flash\_fwd\_splitkv\_kernel} execution time increases monotonically with aggregate KV-cache usage, while compute-bound kernels (\texttt{rms\_norm}, \texttt{elementwise}) remain flat. 
\end{redquestion}

This phenomenon is in line with~\cite{vijaya2025aqua,zhong2024distserve} but from a latency-adversarial perspective. TPOT increase stems from the \textit{memory-bound} nature of the autoregressive decoding phase. In implementations like vLLM with FlashAttention~\cite{flashattention}, the decoding kernel must stream the entire history of KV cache blocks from High Bandwidth Memory (HBM) to on-chip SRAM at every generation step. Essentially, this finding exposes an intriguing architectural insight: while CB is able to thwart latency attacks to certain extent, such logical isolation on the software-level is insufficient to enforce hardware isolation due to memory bandwidth contention. Although such minor contention is too weak to constitute a DoS attack by itself, it creates a deterministic connection between global VRAM usage and individual request latency (indicated by TPOT). This unintended information leak serves as the theoretical foundation, which plays an indispensable part in the proposed attack described next. 

\subsection{Theoretical Analysis}
\label{subsec:theoretical_analysis}

To explain the empirical findings based on theoretical models, we perform analysis to characterize the attack boundaries. Consider a serving system with batch size $B$. Let $r_{\mathrm{adv}}$ denote an adversarial request generating $L_{\mathrm{adv}}$ output tokens (the attacker's goal is $L_{\mathrm{adv}}\rightarrow\infty$), and let $r_n$ be a co-located normal request with expected output length much shorter than $L_{\mathrm{adv}}$. Detailed proofs are provided in Appendix~\ref{appendix_proof}. 

\noindent\textbf{Proposition 1 (CB Logical Isolation).}
In normal cases with no memory pressure, system utilization $\rho < 1 - 1/B$ and batch size $B$, the expected TTFT overhead induced on $r_n$ by $r_{\mathrm{adv}}$ satisfies,
\begin{equation}
\small
\mathbb{E}\!\left[\mathrm{TTFT}_{n}^{\mathrm{attack}}\right]
-
\mathbb{E}\!\left[\mathrm{TTFT}_{n}^{\mathrm{benign}}\right]
\leq
\frac{T_{\mathrm{prefill}}(r_{\mathrm{adv}})}{B},
\label{eq:cb_logical_isolation}
\end{equation}
where $T_{\mathrm{prefill}}(r_{\mathrm{adv}})$ is the prefill latency of $r_{\mathrm{adv}}$. Consequently, the TTFT penalty is independent of $L_{\text{adv}}$ when the system is stable and memory-unconstrained.

Corresponding to Finding \#1, Proposition 1 shows that model-centric latency attacks, which target output length $L_{\text{adv}}$, cannot increase the average TTFT of co-located users beyond a small scheduling overhead, regardless of how many tokens the attacker generates: once $r_{\text{adv}}$ enters the running queue it occupies one out of $B$ slots, but CB continuously replenishes the remaining $B-1$ slots with waiting requests. Hence $r_{\text{adv}}$'s prolonged decode phase does not logically block $r_n$. Corresponding to Finding \#2, since co-located requests still compete for shared HBM bandwidth during KV-cache access, we derive the relation between Inter-Token Latency (ITL) and KV cache utilization.

\noindent\textbf{Proposition 2 (ITL-KV Correlation).}
Let $F_t$ denote the total active KV state referenced across all running requests at decode step $t$. In memory-bound regime, the per-step decode latency $T_t$ is approximately affine in aggregate KV utilization, 
\begin{equation}
\small
T_t \;\approx\; a \cdot \frac{F_t}{BW_{\mathrm{HBM}}} + b, \; \; \text{where} \; \; a = 2 \cdot n_{\text{layers}} \cdot n_{\text{kv\_heads}} \cdot d_{\text{head}} \cdot \mathrm{sizeof}(T).   \label{eq:prop_2}
\end{equation}
$a$ captures the per-token KV memory access cost (both keys and values, across all layers and KV heads, $T$ is data type), and $b$ captures compute-bound overhead independent of context length. 

Proposition 2 explains Finding \#2 and is further validated in Appendix \ref{appendix:side_channel}. The attention kernel (\texttt{flash\_fwd\_splitkv\_kernel}) must stream the full KV-cache for all active contexts from HBM to on-chip SRAM at every generation step. As aggregate KV usage grows, this memory traffic saturates HBM bandwidth, making decode latency a proxy for global memory pressure. This is confirmed by CUDA profiling (Fig.~\ref{fig:kernel_func}): the attention kernel exhibits monotonic growth with KV usage while compute-bound kernels (\texttt{rms\_norm}, \texttt{elementwise\_ops}) remain flat. It lays the foundation for launching the attack in black-box settings discussed next. 

\noindent\textbf{Proposition 3 (FCFS-Induced HOL Blocking).}
Consider a typical FCFS admission queue $\mathcal{Q}_t = (q_1, q_2, \cdots, q_m)$ and $r(q_i)$ as the blocks required to admit request $q_i$ and $F_t$ as the current free block count. Assume $r(q_1) > F_t$, but there exists $q_j$, $j>1$ with $r(q_j) \leq F_t$. If the scheduler does not perform skip-ahead admission, then no request $q_j$ can be admitted until $F_t \geq r(q_1)$ or $q_1$ is removed. The minimum blocking duration for every later request is $W_j \geq \tau(q_1, t) - t$, where $\tau(q_1, t)$ is the first future time when free blocks reach $r(q_1)$. The attacker drives to meet this condition by accumulating,
\begin{equation}
\small
\sum_{r \in \texttt{RUNNING}} \texttt{KV\_blocks}(r) \geq \texttt{total\_blocks} - M_{\mathrm{req}}. \label{eq:prop_3}
\end{equation}
where $\texttt{KV\_blocks}(r)$ is the number of physical blocks allocated to request $r$, $\texttt{total\_blocks}$ is the total GPU pool size and $M_{\mathrm{req}}$ is the blocks required for the HOL waiting request.

\noindent\textbf{Corollary 3 (Preemption Cost).}
When the scheduler preempts a victim request $r_v$ of current sequence length $L_v$, the recovery cost satisfies,
\begin{equation}
\small
T_{\text{recover}}(r_v) = \begin{cases} T_{\text{prefill}}(L_v), & \text{recomputation recovery,} \\ S_{\text{KV}}(L_v) \;/\; BW_{\text{PCIe}}, & \text{swap recovery,} \end{cases}
\end{equation}
where $T_{\text{prefill}} \approx \Theta(L_v^2)$ represents standard transformer where attention computation dominates, and the swapped KV state has size $S_{\text{KV}}(L_v) = 2 L_v \cdot n_{\text{layers}} \cdot n_{\text{kv\_heads}} \cdot d_{\text{head}} \cdot \mathrm{sizeof}(T)$. The attacker's marginal cost to trigger preemption by adding $\delta$ tokens near the saturation boundary is only $O(\delta \cdot \text{price\_per\_token})$, yielding an asymmetry ratio $T_{\text{recover}} / C_{\text{trigger}} \in \Omega(L_v / \delta)$.

Proposition/Corollary 3 together provide a foundation for the proposed F\&S Attack from the cost perspective. The attacker could toggle the system across saturation boundary with minimal tokens, and force the system to repeatedly pay the full recovery cost $T_{\text{recover}}$ for each preemption event, while their own injection cost scales only with $\delta$.

\begin{figure}[htbp]
    \centering
    \includegraphics[width=.82\linewidth]{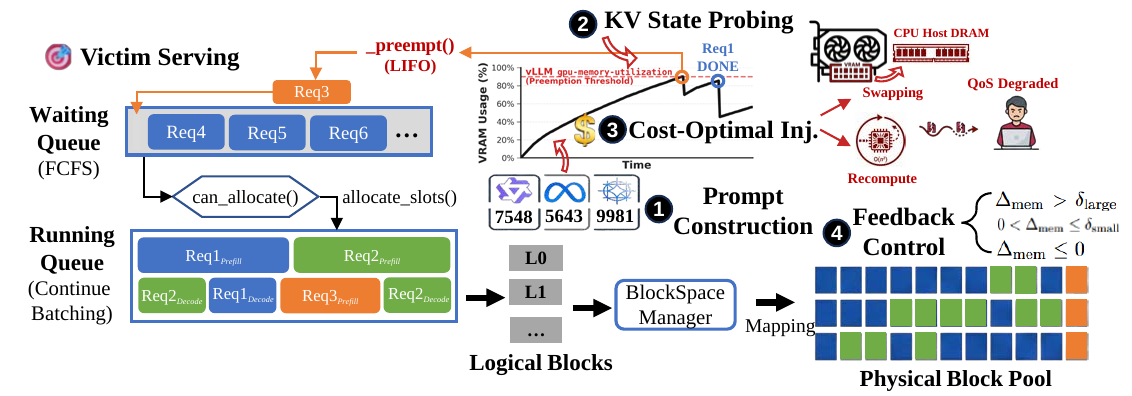}
    \caption{Overview of F\&S attack steps and scheduling states, in which a key is the transition between the FCFS-based \texttt{Waiting Queue} and the continuously-batched \texttt{Running Queue}.}
    \label{fig:serving}
\end{figure}

\section{Attack LLM Serving Framework with Fill and Squeeze}
\label{sec:fill_squeeze_attack}

We propose Fill and Squeeze (F\&S) attack that targets the LLM scheduler's state machine rather than the model's generation process. The key observation is that the LLM serving scheduler only has a few stable states: an IDLE state with sufficient free KV-cache capacity, a \textbf{SATURATED} state where memory-based HOL blocking becomes active, and a \textbf{THRASHING} state where the system oscillates near the saturation boundary due to repeated preemption and recovery. F\&S drives the system from IDLE into THRASHING through: \textbf{Fill}, which rapidly depletes available KV-cache blocks, and \textbf{Squeeze}, which adds just enough load near the boundary to repeatedly trigger costly preemption. E.g., in vLLM~\cite{vllm}, the transition is between the FCFS-based \texttt{WAITING} queue and the continuously batched \texttt{RUNNING} queue. A more detailed interpretation is available in Appendix~\ref{appendix_scheduling_dynamics}. 

The attack operates in four phases: \ding{202} payload-agnostic prompt construction, \ding{203} black-box KV state probing, \ding{204} cost-optimal injection, and \ding{205} adaptive feedback control. These phases jointly enable the attacker to estimate the current KV-cache state, select the cheapest prompt combinations that can cross the saturation boundary, and avoid self-preemption while sustaining disruption.

\ding{202} \textbf{Payload-Agnostic Prompt Construction}. A primary objective of the attacker is to utilize tiered prompts to save attack cost and avoid detection, i.e., $\mathcal{P}=\{\mathcal{P}_{\mathrm{High}},\mathcal{P}_{\mathrm{Mid}},\mathcal{P}_{\mathrm{Low}}\}$ and maximized \textbf{Expansion Ratio} $L_{\mathrm{out}}/L_{\mathrm{in}}$. $\mathcal{P}_{\mathrm{High}}$ fills the KV cache for rapid depletion \texttt{free\_block\_queue} and $\mathcal{P}_{\mathrm{Low}}$ tips an already saturated scheduler into preemption, which is also used for probing. $\mathcal{P}_{\mathrm{Mid}}$ sustains memory pressure during transitional states. Our method supports any pluggable payload (different latency attacks) that can induce long outputs. In this work, we instantiate it with plain-text prompts for black-box transferability~\cite{train_it,feiglin2026benchoverflow} and ExtendAttack~\cite{ExtendAttack} to trigger extended reasoning. 

\ding{203} \textbf{Black-Box KV-Cache Estimation via ITL Probing}. A practical challenge of launching F\&S attack is to know the system KV-cache utilization in real-time. Prior work such as Helix~\cite{helix} assumes privileged access to request output lengths and historical statistics. However, in black-box adversarial settings, the attacker is also an external tenant and lacks direct visibility into the global scheduler state. Let $U_{\mathrm{sys}}$ denote the fraction of used KV-cache blocks over total blocks. Based on Finding \#2 and Proposition~2, we could infer $U_{\mathrm{sys}}$ through the ITL-based side channel induced. Operationally, the attacker periodically dispatches probe requests $p_{\mathrm{probe}}\in\mathcal{P}_{\mathrm{Low}}$ and records their ITL sequence $\mathbf{t}=(t_1,t_2,\ldots,t_k)$ for estimation. 

\begin{figure}[htbp]
    \centering
    \begin{subfigure}[b]{0.365\linewidth}
        \centering
        \includegraphics[width=\linewidth]{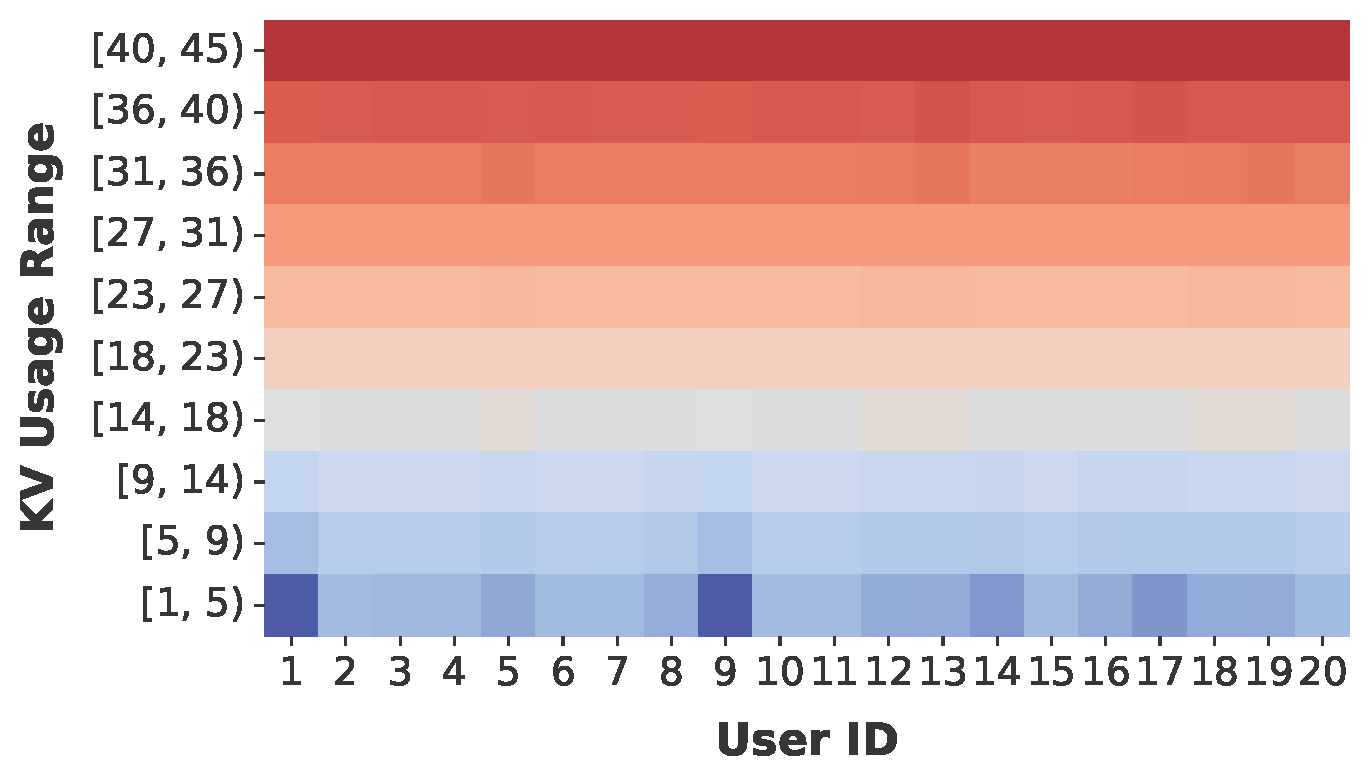}
        \caption{Per-User Trends}
        \label{fig:itl_cache}
    \end{subfigure}
    \begin{subfigure}[b]{0.18\linewidth}
        \centering
        \includegraphics[width=\linewidth]{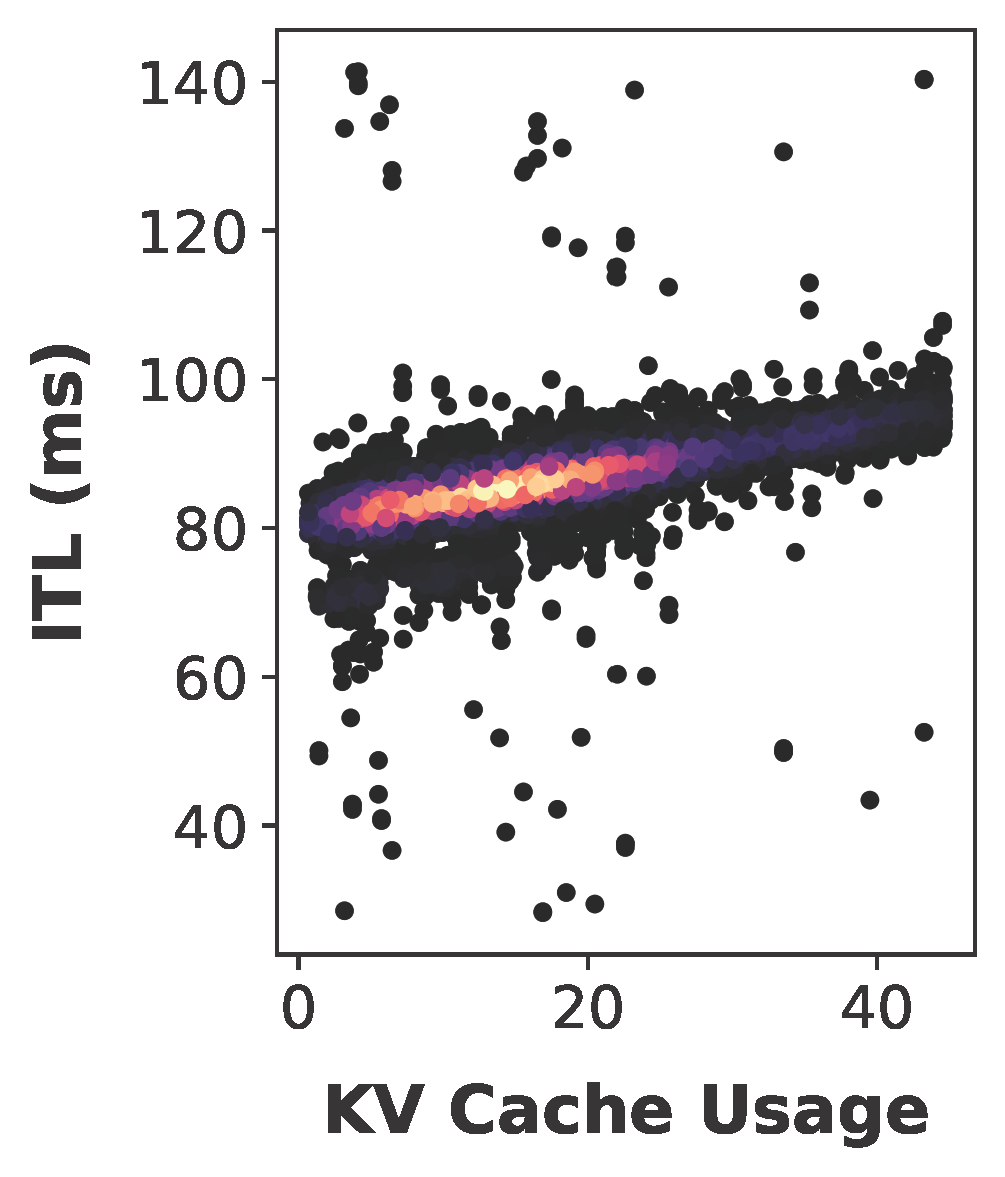}
        \caption{ITL Heatmap}
        \label{fig:heatmap}
    \end{subfigure}
    \begin{subfigure}[b]{0.425\linewidth}
        \centering
        \includegraphics[width=\linewidth]{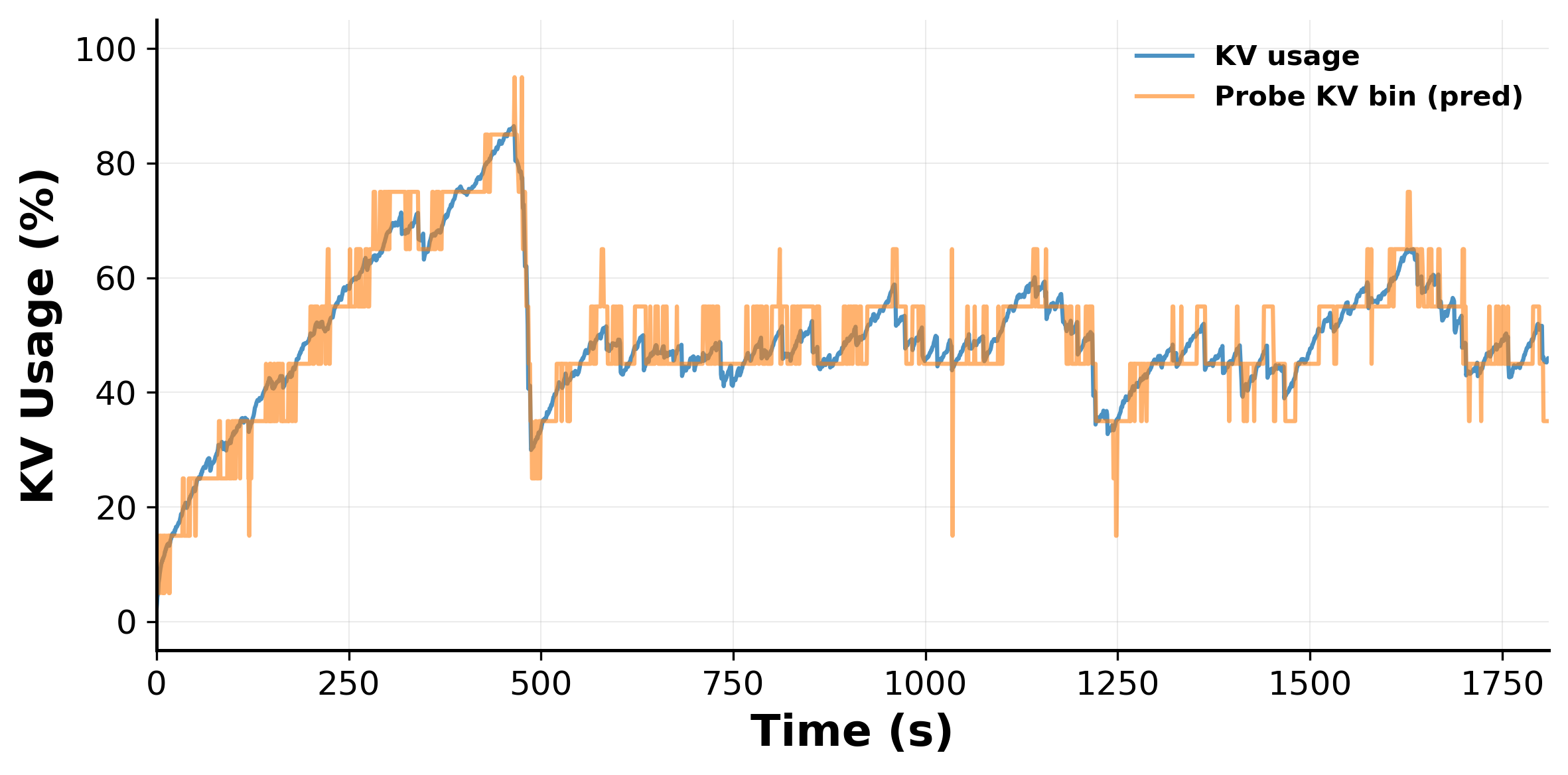}
        \caption{Tracing the KV-Cache Usage Prediction}
        \label{fig:probe}
    \end{subfigure}
    \small{\caption{Black-box estimation of KV usage on Qwen3-8B. (a) Similar latency imposed on all users due to shared memory contention; (b) strong linear correlation between ITL and global KV-cache usage ($U_{\mathrm{sys}}$); (c) trace of KV-cache estimation accuracy using LightGBM.}}
    \label{fig:itl_analysis_combined}
\end{figure}

It turns out that the attacker can use a lightweight classifier, $\mathcal{M}_{\theta}: \mathbb{R}^{k}\rightarrow[0,1]$ to map the ITL sequence to discrete KV-usage bins with high accuracy in real-time. For instance, we can instantiate $\mathcal{M}_{\theta}$ with LightGBM~\cite{ke2017lightgbm}. Compared with sequence models, LightGBM achieves higher accuracy and substantially lower inference latency, which allows the attacker to synchronize injection timing with the scheduler state at sub-second granularity. As shown in Fig.~\ref{fig:probe}, the predicted KV usage consistently tracks the ground truth across both gradual accumulation and abrupt memory release. The low variance in Fig.~\ref{fig:heatmap} further indicates that the probe is robust to stochastic scheduling noise.

\begin{figure*}[t]
\centering
\includegraphics[width=1.0\linewidth]{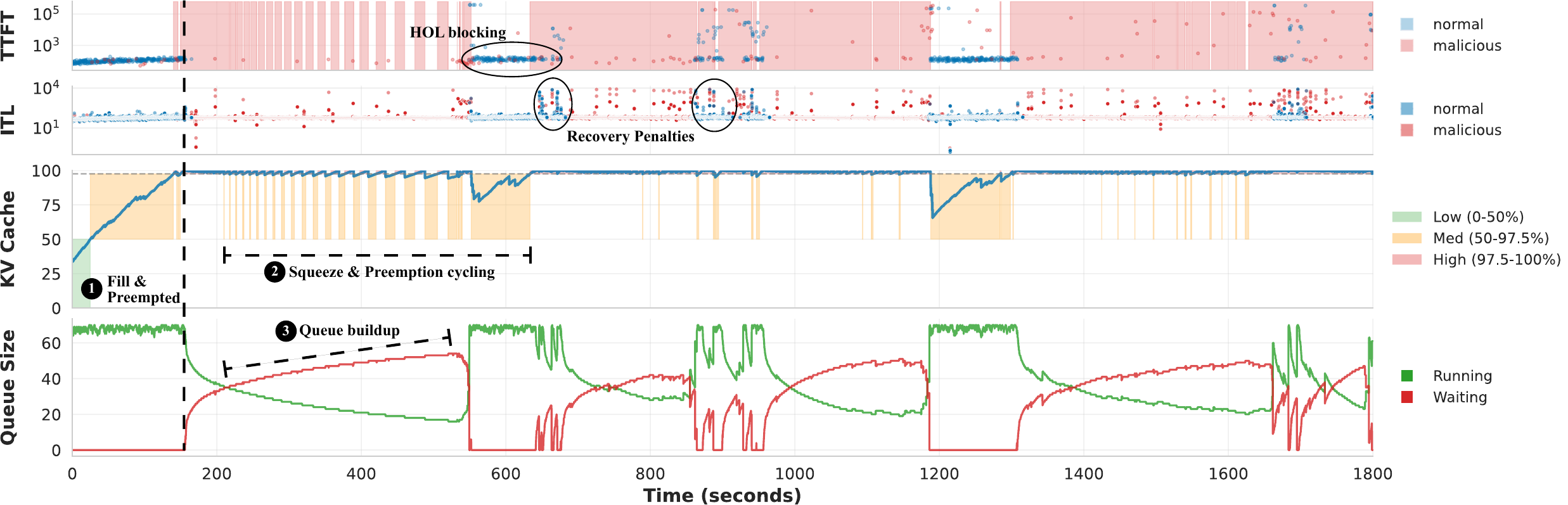}
\caption{\small{Visualization of TTFT, ITL, Global KV-Cache usage and queue size of F\&S attack over time. The Waiting Queue Size (bottom) accumulates as opposed to the Running Queue, which directly correlates with the massive outliers observed in the TTFT Scatter Plot (top right), where user requests are blocked for tens of seconds before processing begins, confirming that the TTFT surge stems from HOL blocking.} }
\vspace{-5pt}
\label{fig:main_fig}
\end{figure*}

\ding{204} \textbf{Cost-Optimal Injection Strategy}. With the set of tiered prompt $\mathcal{P}$ and the estimator $\mathcal{M}_{\theta}$, the attacker selects an injection policy that minimizes time-averaged cost while keeping the system at or above the saturation boundary $C_{\mathrm{sat}}$ identified in Sec.~\ref{appendix_scheduling_dynamics}. Let $p_t\in\mathcal{P}$ denote the prompt selected at time $t$, and let $L_{\mathrm{in}}(p_t)$ and $L_{\mathrm{out}}(p_t)$ denote its input and generated output lengths. We adopt industrial cost model, $\text{Cost} = \alpha \cdot L_{\mathrm{in}} +  \beta \cdot L_{\mathrm{out}}$~\cite{openai2024pricing}. Given unit input and output token prices $\alpha$ and $\beta$ (e.g., $\alpha=\$0.50$, $\beta=\$1.50$ per $10^6$ input/output tokens), F\&S can be formulated as a cost minimization problem,
\begin{equation}
\small
\begin{aligned}
\mathbf{P1}: \quad
\min_{p_1,\ldots,p_T\in\mathcal{P}}
&\quad
\limsup_{T\rightarrow\infty}
\frac{1}{T}
\sum_{t=1}^{T}
\left(
\alpha L_{\mathrm{in}}(p_t)
+
\beta L_{\mathrm{out}}(p_t)
\right) \\
\hspace{-0.3in} \mathrm{s.t.}
&\quad
\mathcal{M}_{\theta}(\mathbf{t}_t)
+
\mathcal{E}(p_t)
\geq
C_{\mathrm{sat}}+\delta,
\quad
\forall t\in[1,T],
\end{aligned}
\label{eq:cost_optimal_injection}
\end{equation}
where $\mathcal{M}_{\theta}(\mathbf{t}_t)$ is the estimated current KV-cache usage from the latest ITL vector, $\mathcal{E}(p_t)$ is the expected memory expansion of prompt $p_t$, and $\delta>0$ is a safety margin ensuring that the system enters the saturation region. To see why our tiered strategy is reasonable, let $\gamma=C_{\mathrm{High}}/C_{\mathrm{Low}}-1$ denotes the cost ratio between high- and low-intensity prompts. There are two cases: 1) If the system is idle, i.e., $U_{\mathrm{sys}} < C_{\mathrm{sat}}-\delta_{\mathrm{large}}$, any feasible policy must fill a large memory gap $\Delta_{\mathrm{mem}}=C_{\mathrm{sat}}-U_{\mathrm{sys}}$. In this regime, $\mathcal{P}_{\mathrm{High}}$ is the cheapest way to add sufficient memory pressure in one shot due to its large expansion ratio; 2) If the system is near saturation, i.e., $U_{\mathrm{sys}}\in[C_{\mathrm{sat}}-\delta_{\mathrm{small}},C_{\mathrm{sat}})$, the constraint in Eq.~\eqref{eq:cost_optimal_injection} can be satisfied using $p\in\mathcal{P}_{\mathrm{Low}}$, since its expected expansion is sufficient to cross the threshold. As shown in our real-world experiments, under ambient traffic, the attack becomes cheaper during peak-load periods as he only needs to bridge the remaining gap to saturation with low-cost prompts.

\ding{205} \textbf{Adaptive Feedback Control}. To implement the optimization strategy, F\&S adopts a closed-loop feedback controller and the attack process is visualized in Fig.\ref{fig:main_fig}. The attacker repeatedly probes the system state, estimates the distance to saturation, and chooses between Fill, Squeeze, and Back-off:

\textbf{Fill.} When $\Delta_{\mathrm{mem}}>\delta_{\mathrm{large}}$, the system is relatively idle. The attacker dispatches high-intensity prompts from $\mathcal{P}_{\mathrm{High}}$ to rapidly consume the \texttt{free\_block\_queue}. This pushes the scheduler toward the HOL-blocking condition. This corresponds to the Fill ramp in Fig.\ref{fig:main_fig}, where the global KV-Cache usage rises rapidly and remains pinned to the High (97.5--100\%) band. 

\textbf{Squeeze.} When $0<\Delta_{\mathrm{mem}}\leq\delta_{\mathrm{small}}$, the system is parked near saturation. The attacker switches to low-intensity prompts from $\mathcal{P}_{\mathrm{Low}}$, which impose just enough memory pressure to cross the saturation threshold. This triggers preemption while minimizing the attacker's own token cost and the system pays a recovery cost through swap or recomputation. In Fig.\ref{fig:main_fig}, it is confirmed by the dense red dots of preempted events in the top subplots and continuous sawtooth patterns on the KV-Cache usage.

\textbf{Adaptive Back-off.} When $\Delta_{\mathrm{mem}}\leq 0$, the system is already beyond the saturation boundary and may preempt the most recently admitted running request. If the attacker's own requests are among the most recent, they risk being selected as victims (\textit{self-preemption}). The back-off mechanism therefore pauses injection, allowing the system to spend cycles recovering a legitimate request before the attacker resumes from a slightly below-saturation state. The overall pseudocode is detailed in Appendix~\ref{app:algorithm}.

\section{Experiments}

\subsection{Threat Model}
\label{sec:threat_model}
We consider an LLM inference service using default vLLM v0.11.2~\cite{vllm}: a shared global KV pool, FCFS admission (no skip-ahead), LIFO preemption, no per-tenant KV quota, and client-visible streaming timestamps, with no input filtering applied~\cite{jain2023baseline,defence,attack_detect}. The attacker is a black-box tenant with only public API access and no knowledge of model parameters or runtime internals; in our usage ``black-box'' means no victim-side access, not that a single frozen estimator transfers across hardware, and the ITL probe of Sec.~\ref{sec:fill_squeeze_attack} is trained on a matching surrogate. Using standard concurrency, the attacker synchronizes long-context queries; when rate limits are enforced per account, multiple identities keep aggregate pressure up, and this still does not bypass entity-, payment-, or global-level enforcement. Unlike prior self-inflicted latency attacks~\cite{engorgio,LoopLLM,ExtendAttack,autodos}, our goal is to degrade service for co-located \emph{benign} users through hardware contention that continuous batching cannot isolate. F\&S targets the setting above and is not a claim of universal bypass: strict entity-level quotas, skip-ahead admission, fair queuing, output buffering, and hard length caps each weaken it in a different way (Sec.~\ref{sec:countermeasure}).

\subsection{Experimental Setup}
\textbf{Serving Environment.} 
We conduct experiments on a server equipped with $8\times$ NVIDIA L40S GPUs (48GB VRAM) powered by vLLM v0.11.2~\cite{vllm}. We use mainstream open-source model families including Qwen3-8B~\cite{qwen3}, Gemma3-12B-it~\cite{gemma} and DeepSeek-R1-Distill-Llama-8B~\cite{deepseek_r1}, and set the \texttt{gpu\_memory\_utilization} to 0.90 to allocate 90\% of the VRAM for KV-Cache and model weights. This setup represents a standard production node and the analysis can be easily extended to large-scale GPU clusters. More setup details are available in Appendix~\ref{appendix_a}. 

\textbf{Workload and Datasets.} 
Following~\cite{fairness,vllm,dynaserve,sun2024llumnix}, we first emulate normal traffic with Poisson distribution for the main results (Sec. \ref{sec:main_results}) and adopt Alpaca~\cite{alpaca} as the primary dataset, since its consistent instruction-following format provides a stable baseline for evaluating core performance metrics. Additionally, we also utilize ShareGPT dataset~\cite{vicuna} for ablation studies. 

\textbf{Baselines.} We benchmark our attack against Engorgio~\cite{engorgio}, LoopLLM~\cite{LoopLLM} and ExtendAttack~\cite{ExtendAttack}. These baselines primarily function as algorithmic complexity attacks that are oblivious of the scheduler's state. For F\&S strategy, we utilize plain-text prompts and ExtendAttack as the underlying methods for cross-model transferability and maximum attack strength, respectively. For fair comparison, we deploy the baselines using a fixed-interval injection. In each slot, the adversarial prompt is randomly sampled from the adversarial prompt collections. This represents the baseline attackers that are not aware of the system state and relevant cost. Note that our study is based on the existing fixed-schedule attacks, rather than adaptive wrappers on the existing schemes. The component ablation in Appendix~\ref{subsec:component_ablation} further demonstrates that merely sending adaptively is still insufficient.

\textbf{Evaluation Metrics}. Attack effectiveness is evaluated via the following metrics. For user-perceived latency, we first examine TTFT. This metric represents the initial wait time and an extended TTFT implies failure to acknowledge user inputs due to queue congestion. Second, we monitor TPOT to assess generation quality. Its degradation represents non-smooth interaction that causes the model to get stuck or hang on generating subsequent tokens. In addition, we also measure the 99th percentile (P99) latency for both TTFT and TPOT. It captures the tail-end user experience, representing the worst-case delay felt by the vast majority of users. To measure the system burden, we track the number of \emph{preemption} (cumulative \texttt{\_preempt()} invocations), which quantifies the frequency of scheduler thrashing and indicates additional overhead of recomputation or memory swap. All main and ablation results are averaged over five independent runs: TTFT/TPOT means and P99 values are first computed over benign requests within each run, preemptions, attack requests and cost are per-run totals, and each reported value is the mean of these run-level statistics. Note that due to heavy-tail latency, certain P99 values (e.g., DeepSeek-R1 with LoopLLM in Table~\ref{tab:main_results}) may appear lower than the mean, which is possible when the tail extends beyond the 99th percentile.

\newcommand{\gc}{\cellcolor{gray!40}}
\newcommand{\gcs}{\cellcolor{gray!30}}
\newcommand{\gct}{\cellcolor{gray!20}}
\begin{table*}[t]
\footnotesize
\centering
\caption{\small{Comparison of attack effectiveness. ↑ and ↓ denote the higher/lower the better, respectively. TTFT/TTFT P99 measured in seconds (\textbf{s}), TPOT/TPOT P99 measured in milliseconds (\textbf{ms}). Each cell is the mean over five independent runs.}}
\label{tab:main_results}
\renewcommand{\arraystretch}{1}
\resizebox{\textwidth}{!}{
\begin{tabular}{clcccccccc}
\toprule
\textbf{Target Model}                            & \textbf{Attack Method}               & \textbf{TTFT (↑)} & \textbf{TTFT P99 (↑)} & \textbf{TPOT (↑)} & \textbf{TPOT P99 (↑)} & \textbf{Preempt\#} & \textbf{Attack Request\# (↓)} & \textbf{Cost (\$) (↓)} \\
\midrule
\multirow{6}{*}{Qwen3-8B}                        & Benign (No Attack)                   & 0.15              & 0.27                  & 34.78             & 49.66                 & 0                  & 0                             & ---                    \\
                                                 & Engorgio                             & 0.12              & 0.24                  & 48.17             & 59.51                 & 0                  & 4613                          & 1.18                   \\
                                                 & \gct LoopLLM                         & \gct 1.80         & \gct 58.05            & \gct 114.12       & \gct 1249.20          & \gct 1104          & \gct 295                      & \gct 0.80              \\
                                                 & ExtendAttack                        & 0.85              & 11.90                 & 80.03             & 474.01                & 538                & 529                           & 0.92                   \\
                                                 & \gc \textbf{F\&S+plain-text}         & \gc 11.05         & \gc 400.13            & \gc 194.45        & \gc 2159.74           & \gc 618            & \gc 180                       & \gc 0.75               \\
                                                 & \gcs \textbf{F\&S+ExtendAttack}     & \gcs 35.95        & \gcs 408.86           & \gcs 181.93       & \gcs 991.19           & \gcs 524           & \gcs 73                       & \gcs 0.42              \\
\midrule
\multirow{6}{*}{Gemma3-12B-it}                   & Benign (No Attack)                   & 0.13              & 0.21                  & 68.00             & 76.28                 & 0                  & 0                             & ---                    \\
                                                 & Engorgio                             & 0.13              & 0.20                  & 68.26             & 77.22                 & 0                  & 2221                          & 0.83                   \\
                                                 & LoopLLM                              & 3.46              & 36.83                 & 126.04            & 193.95                & 3414               & 899                           & 0.71                   \\
                                                 & \gct ExtendAttack                   & \gct 10.55        & \gct 52.71            & \gct 163.50       & \gct 680.00           & \gct 3179          & \gct 684                      & \gct 0.68              \\
                                                 & \gc \textbf{F\&S+plain-text}         & \gc 97.83         & \gc 167.73            & \gc 244.29        & \gc 3544.80           & \gc 1367           & \gc 302                       & \gc 0.67               \\
                                                 & \gcs \textbf{F\&S+ExtendAttack}     & \gcs 104.50       & \gcs 178.36           & \gcs 158.57       & \gcs 1453.84          & \gcs 1380          & \gcs 555                      & \gcs 0.46              \\
\midrule
\multirow{6}{*}{\makecell{DeepSeek-R1-\\Distill-Llama-8B}} & Benign (No Attack)         & 0.08              & 0.14                  & 37.51             & 41.15                 & 0                  & 0                             & ---                    \\
                                                 & Engorgio                             & 0.10              & 0.14                  & 48.86             & 55.80                 & 0                  & 2027                          & 1.17                   \\
                                                 & LoopLLM                              & 1.07              & 0.56                  & 57.37             & 115.30                & 89                 & 992                           & 1.06                   \\
                                                 & \gct ExtendAttack                   & \gct 13.29        & \gct 220.58           & \gct 188.86       & \gct 1021.82          & \gct 773           & \gct 168                      & \gct 0.85              \\
                                                 & \gcs \textbf{F\&S+plain-text}        & \gcs 5.31         & \gcs 110.27           & \gcs 89.52        & \gcs 478.90           & \gcs 1098          & \gcs 321                      & \gcs 0.95              \\
                                                 & \gc \textbf{F\&S+ExtendAttack}      & \gc 16.08         & \gc 284.41            & \gc 157.80        & \gc 1215.48           & \gc 506            & \gc 183                       & \gc 0.86               \\
\bottomrule
\end{tabular}
}
\end{table*}

\subsection{Main Results}
\label{sec:main_results}

First, we analyze the attack effectiveness on a single GPU and scale up to multiple GPUs later. Table~\ref{tab:main_results} demonstrates that our strategy achieves a superior balance between attack efficacy and economic cost compared to the baselines. We can see that the existing scheduler-oblivious attacks rarely drive the system into HOL blocking. In contrast, our state-aware strategy consistently pushes the system to the saturation boundary by maximizing HOL events, yielding catastrophic TTFT inflation. E.g., when paired with plain-text prompts (F\&S+plain-text), we achieve a significant $75\times$ TTFT slowdown on Qwen3-8B (0.15~s $\rightarrow$ 11.05~s), and $742\times$ ($0.13$~s $\rightarrow$ $97.83$~s) on Gemma3-12B-it. Meanwhile, F\&S+ExtendAttack magnifies degradation even further. E.g., on Qwen3-8B, TTFT further increases to $35.95$~s. Both methods carry much lower cost compared to the baselines. 

\setlength{\intextsep}{0pt}       
\begin{wraptable}{r}{0.55\textwidth} 
    \centering
    \caption{\small{Workload sensitivity of F\&S attack strategy. $\Delta$ indicates degradation ratio. TTFT/TPOT in s/ms.}}
    \label{tab:workload_ablation}
    \footnotesize 
    \setlength{\tabcolsep}{3pt}
    \renewcommand{\arraystretch}{1.1}
    \begin{tabular}{clcccc}
        \toprule
        \textbf{Model} & \textbf{Workload} & \textbf{TTFT} & \textbf{$\Delta$T} & \textbf{TPOT} & \textbf{$\Delta$P} \\
        \midrule
        \multirow{2}{*}{Qwen3-8B}
         & Alpaca    & 11.05 & 75.6$\times$ & 194 & 5.6$\times$ \\
         & ShareGPT  & 2.57  & 41.0$\times$ & 254 & 7.3$\times$ \\
        \midrule
        \multirow{2}{*}{Gemma3-12B-it}
         & Alpaca    & 97.83 & 742.8$\times$ & 244 & 3.6$\times$ \\
         & ShareGPT  & 65.89 & 720.6$\times$ & 359 & 5.3$\times$ \\
        \midrule
        \multirow{2}{*}{\makecell[l]{DeepSeek-R1-\\D-Llama-8B}}
         & Alpaca    & 5.31  & 68.2$\times$ & 90 & 2.4$\times$ \\
         & ShareGPT  & 5.28  & 67.8$\times$ & 80 & 2.1$\times$ \\
        \bottomrule
    \end{tabular}
\end{wraptable}

\subsection{Ablation Study}
\label{sec:ablation}

\begin{figure}[t]
\centering
\includegraphics[width=\linewidth]{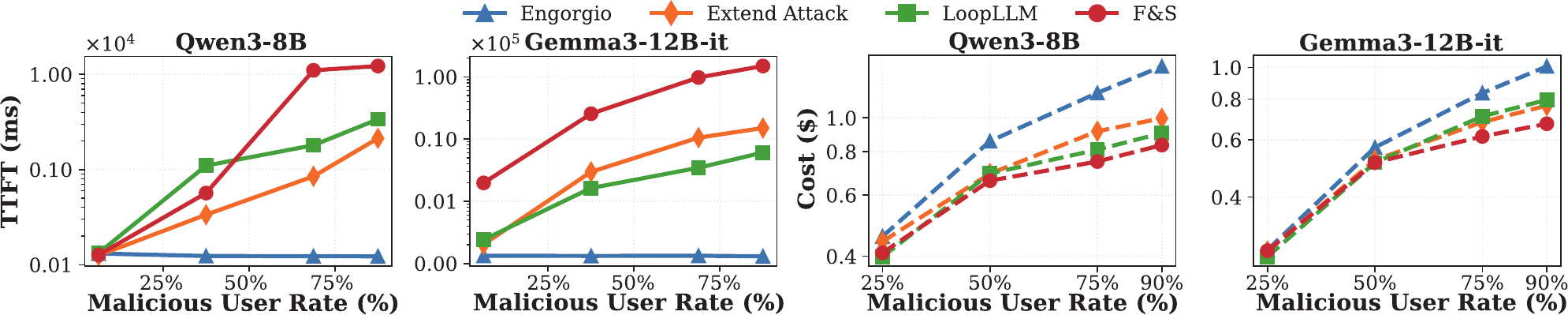}
\caption{\small{Impact of malicious user rate on attack efficacy and cost across different model architectures.}}
\label{fig:ablation_rate}
\end{figure}

\textbf{Attack Intensity}. To quantify how attack efficacy and economic cost evolve with attack intensity, we vary the percentage of malicious users while keeping the total request rate constant. Fig.~\ref{fig:ablation_rate} reveals a clear separation: across model architectures, baseline methods exhibit a gradual TTFT increase as the malicious ratio grows, whereas our method causes a faster performance degradation. Interestingly, the jump of TTFT does not necessitate an equivalent increase of attack cost. Cost evaluations on Fig.~\ref{fig:ablation_rate} (right) reveal that baseline methods incur costs that scale linearly with attack intensity, because \textit{F\&S} employs low-intensity prompts during ``Squeeze'', thereby continuously strengthening the attack at a much lower monetary cost. Empirical results on Gemma3-12B-it show that \textit{F\&S} achieves a 733.9$\times$ relative QoS degradation at 73.3\% of the baseline cost. 

\textbf{Impact of Workload Distributions}. We further evaluate the impact of workload characteristics on F\&S attack. Table~\ref{tab:workload_ablation} demonstrates the results across two contrastive profiles: {Alpaca} (short-context instruction following)~\cite{alpaca} and {ShareGPT} (long-context dialogue)~\cite{vicuna}. It reveals an intriguing phenomenon that challenges the prevailing assumption that long-context traffic is more vulnerable. Notably, the most acute \emph{TTFT} degradation frequently occurs under short-context workloads. E.g., on Qwen3-8B, Alpaca exhibits a {$75.6\times$} TTFT degradation ($11.05$ s), far exceeding the $41.0\times$ observed on ShareGPT ($2.57$ s), and Gemma3-12B-it follows a similar trajectory. In contrast, DeepSeek-R1 remains relatively unchanged across both distributions (68.2$\times$ vs. 67.8$\times$), suggesting an architectural stability to workload-induced variance.

Further, TTFT and TPOT exhibit divergent sensitivities to these workload shifts. While Alpaca triggers higher TTFT inflation on Qwen3-8B and Gemma3-12B-it, ShareGPT consistently incurs more severe TPOT degradation (e.g., 7.3$\times$ vs.\ 5.6$\times$ on Qwen3-8B). This result aligns with the two-stage design of F\&S: \textit{Fill} acts as an admission bottleneck that dominates TTFT by inducing queuing delays, whereas the sustained decoding phase of long-context dialogues amplifies the interference once the system is saturated (occupying TPOT). These findings highlight a critical risk in production environments that high-throughput, lightweight workloads can suffer catastrophic responsiveness stalls or even timeouts, while long-context sessions experience persistent throughput erosion. Hence, designers of the scheduling logic should consider diverse request distributions to be prepared for adversarial, extreme situations. 

\subsection{Realistic Production Scenarios}
\label{sec:scenarios}

\begin{figure}[htbp]
    \centering
    \includegraphics[width=.9\linewidth]{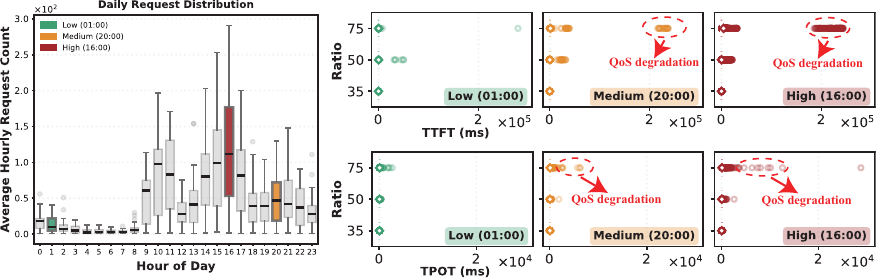}
    \caption{\small{Attack impact on normal users under real-world traffic patterns. Left: Hourly request distribution from BurstGPT traces with three representative traffic intensities. Right: TTFT and TPOT distributions for normal users under varying malicious ratios.}}
    \label{fig:real_world}
\end{figure}

\vspace{10pt}

\begin{figure}[htbp]
    \centering
    \subfloat[\small{Multi-GPU impact}]{
        \includegraphics[width=0.46\linewidth]{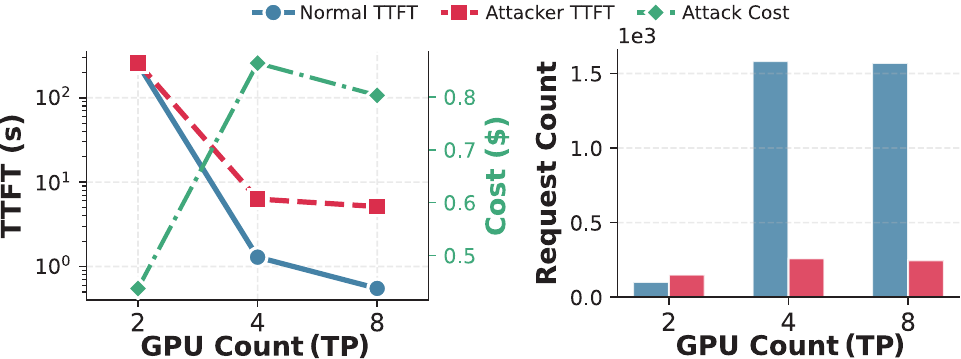}
        \label{fig:sub_a}
    }
    \subfloat[\small{User load impact}]{
        \includegraphics[width=0.43\linewidth]{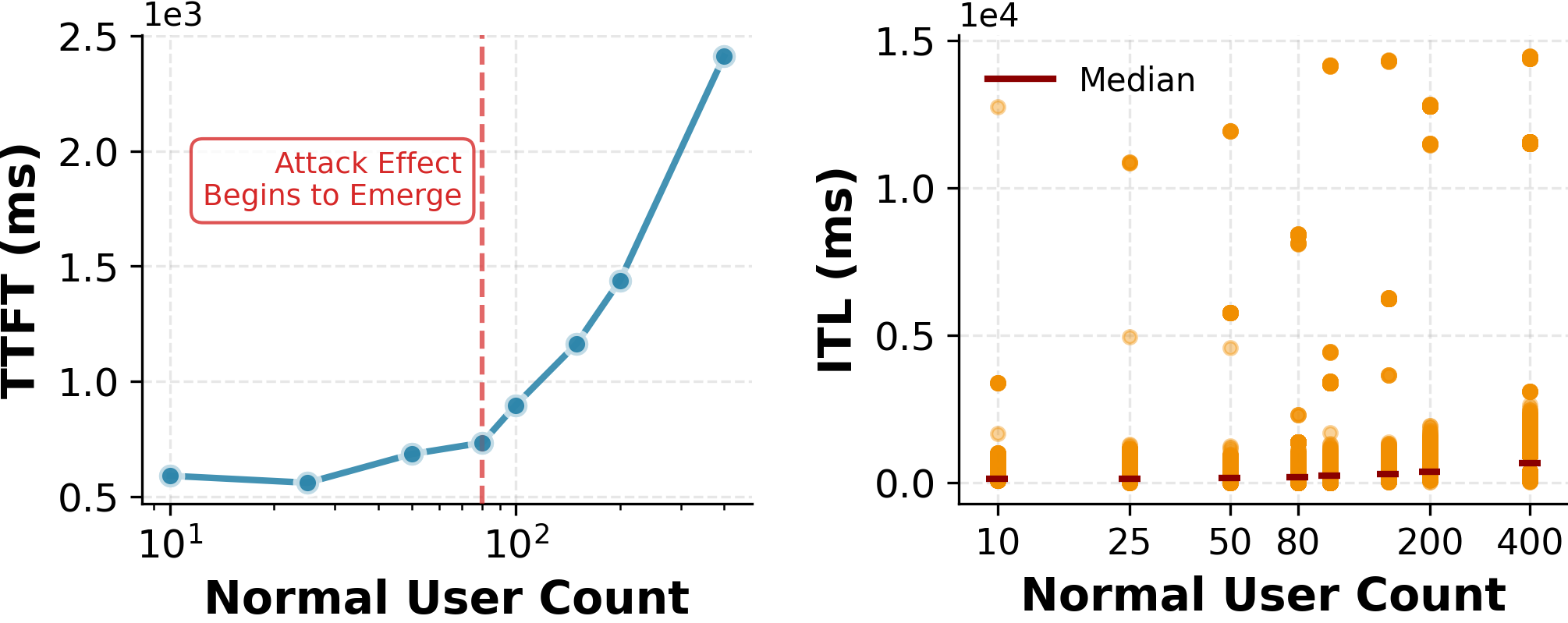}
        \label{fig:sub_b}
    }
    \caption{\small{Multi-GPU scalability analysis under varying hardware resources and load configurations.}}
    \label{fig:multi_gpu}
\end{figure}

\textbf{Real-World Workload}. To evaluate the F\&S attack under realistic deployment conditions, we emulate real-world traffic patterns based on the BurstGPT dataset~\cite{burstgpt} (10.31M real traces). We categorize the workload into three representative time windows as highlighted by different colors on Fig.~\ref{fig:real_world} (Left). The results indicate that the attacker could take advantage of ambient traffic patterns to reduce cost substantially, and ambient traffic acts as a \emph{force multiplier}. By naturally consuming the KV-Cache headroom, the normal, benign requests lower the saturation barrier that allows the attacker to succeed more easily with 40\% less cost. Visualized in Fig.~\ref{fig:real_world} (Right), this results in volatility for both TTFT and TPOT during high traffic.

\textbf{Multi-GPU Tensor Parallelism}. Intuitively, a natural defense against resource exhaustion is over-provisioning. By scaling up resources, service providers aim to raise the bar for the attackers. Thus, to evaluate how F\&S scales against horizontal scaling, we fix the attack strengths and examine its performance across varying degrees of Tensor Parallelism (TP) on the $8\times$ NVIDIA L40S server. As expected, expanding from TP2 to TP4 significantly mitigates user-perceived latency (TTFT) by raising the memory saturation barrier shown in Fig.~\ref{fig:multi_gpu}(a). The expanded KV-Cache capacity prevents the attacker from successfully triggering the \textit{Squeeze} phase (repetitive preemption), forcing the adversary to remain stuck in the capital-intensive ``Fill'' phase, continuously dispatching expensive High-Intensity prompts ($P_{High}$) without inducing thrashing. TP8 provides ample resources that allow benign users to aggressively occupy scheduling slots, effectively starving the attacker of admission and neutralizing the threat at a lower total cost. Other vLLM parameters primarily shift the attack cost profile rather than its feasibility. E.g., \texttt{block\_size} controls allocation granularity near saturation; preemption mode (recomputation vs. swap) and \texttt{swap\_space} determine the per-preemption recovery cost, and the scheduling policy dictates whether FCFS-HOL exists. A detailed parameter-to-mechanism mapping is provided in Appendix~\ref{subsec:vllm_config_mapping}.

\setlength{\intextsep}{0pt}       
\begin{wraptable}{r}{5.8cm} 
    \centering
    \caption{\small{Attack effectiveness and cost analysis under reasoning.}}
    \label{tab:ablation}
    \scriptsize 
    \setlength{\tabcolsep}{2pt} 
    \renewcommand{\arraystretch}{1.1}
    \begin{tabular}{clccc}
        \toprule
        \textbf{Model} & \textbf{Method} & \textbf{TTFT} & \textbf{TPOT} & \textbf{Cost(\$)} \\
        \midrule
        \multirow{5}{*}{\rotatebox{90}{Qwen3-8B}}
        & Engorgio          & 0.10  & 45.9   & 1.23 \\
        & ExtendAttack      & 20.11 & 191.0  & 0.75 \\
        & LoopLLM           & 2.11  & 64.7   & 0.95 \\
        & \gcs \textbf{F\&S+plain-text} & \gcs 21.32 & \gcs 135.2 & \gcs 0.78 \\
        & \gc \textbf{F\&S+ExtendAttack}    & \gc 27.65  & \gc 196.99 & \gc 0.68 \\
        \midrule
        \multirow{5}{*}{\rotatebox{90}{DeepSeek*}}
        & Engorgio          & 0.10  & 50.19  & 1.13 \\
        & ExtendAttack      & 10.99 & 190.96 & 0.70 \\
        & LoopLLM           & 0.16  & 58.02  & 1.00 \\
        & \gcs \textbf{F\&S+plain-text} & \gcs 1.10  & \gcs 67.47  & \gcs 0.93 \\
        & \gc \textbf{F\&S+ExtendAttack}    & \gc 13.24 & \gc 142.34 & \gc 0.74 \\
        \bottomrule
    \end{tabular}
\end{wraptable}

\textbf{Generalization to Reasoning}. Table~\ref{tab:ablation} reports the attack efficacy on Qwen3-8B and DeepSeek-R1-Distill-Llama-8B with reasoning tasks. While traditional baselines like \textit{Engorgio} incur high costs (\$1.23) with negligible impact (TTFT $\approx$ 0.10s), our results highlight the composability of the F\&S strategy. Specifically, combining F\&S with \textit{ExtendAttack} consistently amplifies performance degradation across both architectures. On Qwen3-8B, this hybrid approach pushes TTFT to 27.65 s and TPOT to 196.99 ms, surpassing \textit{ExtendAttack} alone (20.11 s / 191.0 ms) while reducing the financial cost to \$0.68. This enhancement is equally evident on DeepSeek-R1, where F\&S boosts the TTFT of \textit{ExtendAttack} from 10.99 s to 13.24 s. 

\section{Conclusion and Future Work}
\label{sec:countermeasure}
In this paper, we show that existing model-centric latency attacks only impose minor performance impact on modern serving systems, because continuous batching isolates co-located requests at the iteration level. We therefore target the scheduler instead. By exploiting the ITL side channel and tiered prompts, F\&S pushes the shared KV cache into saturation and repeated preemption, and inflicts much higher latency on co-located users at lower attack cost. 

On the other hand, simple defense such as per-account rate or concurrency limits only slow one Fill instance when requests are spread across identities; a KV quota enforced per entity below the saturation threshold keeps that entity from inducing Fill, and hard output caps take away the retained KV that Fill needs. Skip-ahead admission, fair queuing~\cite{fairness}, and priority scheduling~\cite{ikram2025ascendra} remove most FCFS head-of-line blocking, though aggressive fairness can starve legitimate long-context users, and KV held by running requests still sustains the TPOT slowdown from Squeeze. If the provider buffers streaming output, the ITL probe loses its signal and F\&S falls back to a less efficient open-loop attack. Perplexity filters~\cite{attack_detect,phute2023llm} catch optimization-based prompts, but the plain-text prompts used by F\&S read as ordinary text and pass through. Thus, we leave a much larger scale evaluation of F\&S-type of resource consumption attack and defense to the future work. 

\begin{ack}
We thank the anonymous reviewers and the area chair for their detailed feedback, in particular the questions on threat-model scope, provider-side safeguards, and statistical reporting, which led to the added component ablation, production prober evaluation, and countermeasure discussion. This work is supported by National Science Foundation of China under grants 62576310, Zhejiang Provincial National Science Foundation of China under Grant No. LZ25F020007, ICT2026A20 and Fong Shu Fook Tong Research Innovation Program, Zhejiang University Education Foundation.
\end{ack}

\bibliographystyle{plain}
\bibliography{reference}

\appendix
\clearpage

\section{Broader Impacts and Responsible Disclosure}

\textbf{Responsible Disclosure}. This work strictly adheres to established ethical standards in AI \& Cyber research. All experiments are conducted in controlled and isolated environments, and no attacks are launched against public LLM services to avoid potential disruption to real-world users. Our evaluation relies exclusively on publicly datasets without collecting or processing any private user data, and all adversarial inputs are reviewed to exclude harmful/offensive content. As the proposed method has been timely archived by \url{https://www.promptfoo.dev/}, code artifacts consist of benchmarking scripts and evaluation harnesses are available via the anonymous link. We would double check for ethical issues before making them public, but not as attack tools or prompt arsenals.

\textbf{Societal Impact}. This work exposes a previously understudied class of system-level attack surface in LLM serving frameworks. Understanding these attack surfaces enables service providers to design more robust schedulers, admission controls, and per-tenant isolation mechanisms. Defense mechanisms discussed in Appendix~\ref{sec:appendix_countermeasures} provide an actionable guidance for practitioners. While this paper proposes a new system-level attack, our ultimate goal is defensive. We hope it serves as a wake-up call to service providers as current systems designers often treat latency degradation as a \emph{performance issue} rather than a \emph{security threat}. It is essential to understand that DoS attack in LLMs carries equivalent damage potential to conventional network DDoS attacks with asymmetric monetary cost from the attacker and service provider, especially the operational cost and carbon emissions from datacenters are becoming pressing issues. The second note is that this work provides a quantitative framework for service providers to ``red team'' their own infrastructure as a stress test, such that they could align attacker’s cost vs. system degradation. This would allow providers to identify essential boundaries from a wide array of system parameters, and help providers design better admission control policies that remain robust in a competitive and fluid environment.

\section{Technical Appendices and Supplementary Material}

\subsection{Threat Model}\label{appendix:threat_model}

The full threat model, including the five deployment assumptions (shared global KV-cache pool; FCFS admission; preemption enabled; no per-tenant KV isolation; visible streaming per-token timestamps), the attacker's capabilities and limits, and the explicit scope-and-boundary statement of our claims, is stated in the main text (Section~\ref{sec:threat_model}). We do not repeat it here to avoid duplication; the discussion in Appendix~\ref{sec:appendix_countermeasures} on countermeasures uses the same terminology and boundary conditions.

\subsection{Theoretical Analysis}
\label{appendix_proof}

Our objective here is to identify \emph{sufficient conditions} under which latency amplification can emerge in standard continuous-batching LLM serving with default settings, but not to prove or exhaust that all serving frameworks with various customized scheduling policies. 

\noindent\textbf{Proposition 1 (CB Logical Isolation).}
In normal cases with no memory pressure, system utilization $\rho < 1 - 1/B$ and batch size $B$, the expected TTFT overhead induced on $r_n$ by $r_{\mathrm{adv}}$ satisfies, where $T_{\mathrm{prefill}}(r_{\mathrm{adv}})$ is the prefill latency of $r_{\mathrm{adv}}$. Consequently, the TTFT penalty is independent of $L_{\text{adv}}$ when the system is stable and memory-unconstrained.

\noindent\textit{Proof.}
Under CB, the scheduler assigns requests to iteration slots. Once $r_{\mathrm{adv}}$ enters the running queue, it occupies one out of $B$ slots, leaving the remaining $B-1$ slots available for normal requests. Thus, $r_{\mathrm{adv}}$ can delay normal requests only if it occupies the slot that would otherwise admit normal requests. Since CB continuously admits new requests whenever running requests complete, the additional waiting time of normal requests is bounded by the fraction of scheduling capacity consumed by $r_{\mathrm{adv}}$, namely approximately $1/B$ of the prefill cost. If $\Delta_{\mathrm{slot}}$ denotes the duration of one scheduling iteration, then the expected waiting time satisfies,
\begin{equation}
\small
\mathbb{E}[W_n]
\approx
\mathbb{E}\!\left[
\frac{\texttt{slots\_before}(r_n)}{B}\,\Delta_{\mathrm{slot}}
\right],
\end{equation}
which yields Eq.~\eqref{eq:cb_logical_isolation}. For typical production settings, e.g., $B\geq 16$ and $T_{\mathrm{prefill}}\approx 100$ ms, the overhead is below $6.25$ ms per adversarial request. The key insight is that the decoding phase of $r_{\mathrm{adv}}$, even if extremely long, does not by itself block normal requests, because CB continuously replenishes freed slots with waiting requests. \hfill$\square$

\noindent\textbf{Proposition 2 (ITL-KV Correlation).}
Let $F_t$ denote the total active KV state referenced across all running requests at decode step $t$. In memory-bound regime, the per-step decode latency $T_t$ is approximately affine in aggregate KV utilization, $\alpha$ captures the per-token KV memory access cost (both keys and values, across all layers and KV heads, $T$ is data type), and $\beta$ captures compute-bound overhead independent of context length.

\noindent\textit{Proof.}
During decoding, the attention kernel streams KV-cache blocks from HBM to on-chip SRAM. For a single request, the attention time scales linearly with context length:
\begin{equation}
\small
T_{\mathrm{attn}}(r_i)
\approx
\frac{
N_{\mathrm{ctx}}(r_i)\,d_{\mathrm{model}}\,\texttt{sizeof(T)}
}{
\mathrm{BW}_{\mathrm{HBM}}
}.
\label{eq:single_request_attn}
\end{equation}
With $N$ concurrent requests, their attention kernels contend for the same HBM bandwidth. Under the \emph{Roofline Model}~\cite{williams2009roofline}, the effective service time grows with aggregate memory demand. Therefore, the attention component can be rewritten as a function of the total active context length, yielding Eq.~\eqref{eq:prop_2}. Compute-bound operations, such as normalization and elementwise kernels, contribute the approximately constant term $\beta$, while the attention component grows with KV-cache utilization. \hfill$\square$

In addition, CUDA profiling shows that attention decode kernels increase with KV-cache usage, whereas compute-bound kernels remain comparatively stable. The measured $R^2$ values further confirm a strong ITL--KV correlation under production-relevant concurrency levels. 


\noindent\textbf{Proposition 3 (FCFS-Induced HOL Blocking).}
Consider a typical FCFS admission queue $\mathcal{Q}_t = (q_1, q_2, \cdots, q_m)$ and $r(q_i)$ as the blocks required to admit request $q_i$ and $F_t$ as the current free block count. Assume $r(q_1) > F_t$, but there exists $q_j$, $j>1$ with $r(q_j) \leq F_t$. If the scheduler does not perform skip-ahead admission, then no request $q_j$ can be admitted until $F_t \geq r(q_1)$ or $q_1$ is removed. The minimum blocking duration for every later request is $W_j \geq \tau(q_1, t) - t$, where $\tau(q_1, t)$ is the first future time when free blocks reach $r(q_1)$. The attacker drives to meet this condition by accumulating,
where $\texttt{KV\_blocks}(r)$ is the number of physical blocks allocated to request $r$, $\texttt{total\_blocks}$ is the total GPU pool size and $M_{\mathrm{req}}$ is the blocks required for the HOL waiting request.

\noindent\textit{Proof.} The scheduler iterates through the \texttt{WAITING} queue in FCFS order and invokes \texttt{can\_allocate()} for each request. This function returns \texttt{False} when \texttt{free\_block\_queue.num\_free\_blocks} $< M_{\mathrm{req}}$. Upon the first allocation failure, the scheduler terminates the admission loop rather than skipping smaller requests, thereby preserving FCFS fairness. Hence, once the HOL requests cannot be allocated, all subsequent waiting requests are also prevented from entering the running queue, which establishes the HOL-blocking condition. The stall persists until the running queue frees at least $M_{\mathrm{req}}$ KV blocks. Since running requests release KV blocks only upon completion, recomputation-based recovery, or explicit preemption, the minimum block-freeing rate is bounded by decode throughput, yielding Eq.~\eqref{eq:prop_3}. \hfill$\square$

\noindent\textbf{Corollary 1 (Preemption Cost).}
When the scheduler preempts a victim request $r_v$ of current sequence length $L_v$, the recovery cost satisfies,
\begin{equation}
\small
T_{\text{recover}}(r_v) = \begin{cases} T_{\text{prefill}}(L_v), & \text{recomputation recovery,} \\ S_{\text{KV}}(L_v) \;/\; BW_{\text{PCIe}}, & \text{swap recovery,} \end{cases}
\end{equation}
where $T_{\text{prefill}} \approx \Theta(L_v^2)$ represents standard transformer where attention computation dominates, and the swapped KV state has size $S_{\text{KV}}(L_v) = 2 L_v \cdot n_{\text{layers}} \cdot n_{\text{kv\_heads}} \cdot d_{\text{head}} \cdot \mathrm{sizeof}(T)$. The attacker's marginal cost to trigger preemption by adding $\delta$ tokens near the saturation boundary is only $O(\delta \cdot \text{price\_per\_token})$, yielding an asymmetry ratio $T_{\text{recover}} / C_{\text{trigger}} \in \Omega(L_v / \delta)$.

\noindent\textit{Proof.} For recomputation-based recovery, the scheduler discards the victim's KV cache and replays the forward pass over the full accumulated sequence, including both the prompt and previously generated tokens. Transformer forward complexity over $L_v$ tokens is $O(L_v d^2 + L_v^2 d)$, where the quadratic attention term dominates for long contexts, yielding $\Theta(L_v^2)$ cost. For swap-based recovery, the KV cache stores keys and values for each layer and each token; transferring this state over PCIe gives latency linear in $L_v$, namely $\Omega(L_v/\mathrm{BW}_{\mathrm{PCIe}})$. The attacker, however, only needs to push aggregate KV usage beyond the preemption threshold. Near the saturation boundary, this may require only one additional KV block, so the triggering cost scales with $\delta$, whereas the victim's recovery cost scales with $L_v$. \hfill$\square$

Note that in practice, the wall-clock recovery time can be reduced via FlashAttention, chunked prefill and prefix caching. The analysis above provides an upper bound of the recovery cost.

\subsection{vLLM Scheduler Basics} \label{appendix_scheduling_dynamics}

\subsubsection{KV-Cache Saturation}
The physical limit of GPU VRAM is a hard constraint in LLM serving. In vLLM, memory management is orchestrated by the \texttt{BlockSpaceManager}\footnote{Source code references are based on the official vLLM repository: \url{https://github.com/vllm-project/vllm}.}, which maps logical KV blocks to physical GPU memory. 

To manage concurrency, the scheduler maintains a \texttt{WAITING} queue for pending requests and a \texttt{RUNNING} queue for those currently executing on the GPU. During the autoregressive decoding phase, the memory footprint of every active request grows monotonically. For each inference iteration, the \texttt{Scheduler} assigns new physical blocks from the global \texttt{free\_block\_queue}. 
Unlike computational slots that are ephemeral and reset after every step, the KV-Cache blocks allocated to running sequences are persistent. As requests in the \texttt{RUNNING} queue progress through the decoding phase, they continuously consume the remaining capacity in the \texttt{free\_block\_queue}. An adversary can take advantage of this by exploiting long sequences to quickly deplete the memory resources and drive the system toward a saturation point. 


When this physical limit is reached, the scheduler's admission logic creates a bottleneck. Following a strict First-Come-First-Served (FCFS) policy, the scheduler iterates through the \texttt{WAITING} queue and invokes \texttt{can\_allocate()} for the request at the head of the line. If this check fails due to memory exhaustion, scheduling terminates immediately to preserve request ordering and fairness, which we term as \emph{Memory-Based HOL Blocking}. This blockage creates a DoS situation even if the computational capacity is still available, i.e., global \texttt{num\_batched\_tokens} budget used by \texttt{chunked\_prefill} is sufficient to accommodate subsequent, smaller requests. But due to physical block unavailability, legitimate short requests are stalled indefinitely.

\subsubsection{Preemption and Recovery Penalties}

The vulnerability emerges when the system attempts to recover from this saturation. When the cumulative memory demand of \texttt{RUNNING} requests surpasses the physical capacity, the scheduler must reclaim resources to prevent a deadlock. It triggers a deterministic preemption mechanism by invoking \texttt{\_preempt()} to free up space. To minimize the impact on long-running tasks, vLLM typically enforces a Last-In-First-Out (LIFO) policy and selects the most recently scheduled request ($r_{victim}$) to evict. For fairness, the preempted victim is not discarded but prioritized by being prepended to the front of the \texttt{WAITING} queue. Depending on different implementations, vLLM operates under one of the following principles for recovery: 
\begin{itemize}
\item[\ding{70}] \textbf{Swapping (I/O Bottleneck):} Common in legacy architectures (e.g., vLLMv0), memory swap offloads the victim's KV blocks from high-speed GPU VRAM to CPU host memory. Recovery requires migrating these blocks back over the bandwidth-constrained PCIe bus. 
\item[\ding{70}] \textbf{Recomputation (Compute Bottleneck):} In unified scheduler architectures (e.g., vLLMv1), the system prioritizes implementation simplicity by discarding the victim's KV-Cache entirely. To resume execution, the engine must re-process the prompt and all previously generated tokens from scratch. This shifts the workload from the linear-time decoding back to the quadratic-time prefilling.
\end{itemize}
While preemption is intended to ensure stability, it exposes an asymmetry facing adversaries. The attacker can trigger a preemption with a marginal increase in memory usage, yet the system must pay a disproportionately high cost to recover the victim request later. By toggling across the saturation boundary, the attacker can induce the system into back-and-forth preemption thrashing loop that continuously imposes pressure on the memory hierarchy. 



\subsection{Attack Prompt Construction}

Here, we provide more detail and analysis of the attack cost and how we finally reach the compositional prompt sets.

\textbf{Attack Cost Analysis}. While existing literature has proposed a wide range of adversarial techniques~\cite{engorgio, LoopLLM, ExtendAttack, autodos}, their economic viability remains underexplored. In production environments, the usage is metered by pay-per-use billing so monetary expenditure plays an essential role for launching an attack. To evaluate different attack strategies, we formalize the cost structures into three paradigms: \ding{202} white-box attacks that craft adversarial prompts on a surrogate model and transfer prompts to the target model~\cite{engorgio, LoopLLM}; \ding{203} black-box attacks that repetitively query an API to craft adversarial prompts~\cite{ExtendAttack, autodos}; \ding{204} plain-text prompts that utilize deterministic logic and instruction-follow capability of LLM that are natural and non-adversarial~\cite{feiglin2026benchoverflow}. Their monetary cost can be quantified as,
{\small
\begin{equation}
C =
\begin{cases}
T_{\text{opt}} \cdot P_{\text{avg}} \cdot p_e, & \text{(White-Box)} \\
\sum_{j=1}^{N} \left( H_{i, j} \cdot p_i + H_{o, j} \cdot p_o \right), & \text{(Black-Box)} \\
H_i \cdot p_i + H_o \cdot p_o. & \text{(Plain-text)}
\end{cases}
\label{eq:cost_analysis}
\end{equation}
}
\begin{wrapfigure}{r}{0.6\textwidth} 
    \centering
    \captionsetup[subfigure]{font=scriptsize}
    
    \begin{subfigure}[b]{0.45\linewidth}
        \centering
        \includegraphics[width=\linewidth]{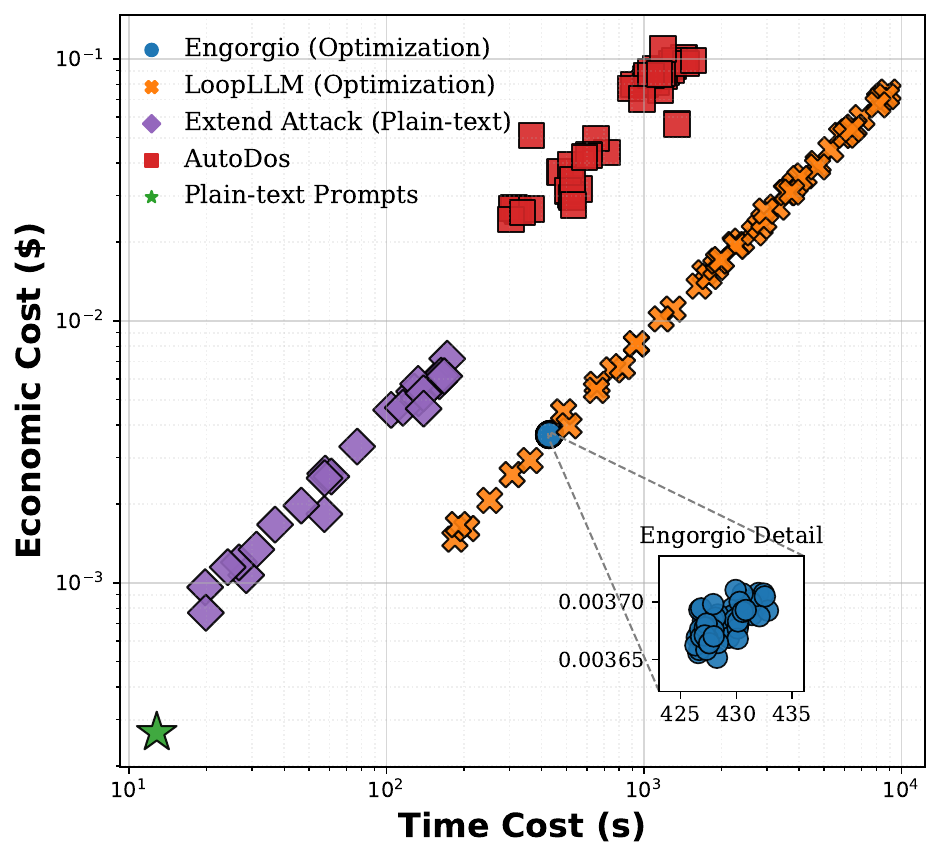}
        \caption{Economic vs. Time Cost}
        \label{fig:cost_scatter}
    \end{subfigure}
    \hfill 
    \begin{minipage}[b]{0.45\linewidth}
        \centering
        \begin{subfigure}{\linewidth}
            \centering
            \includegraphics[width=\linewidth]{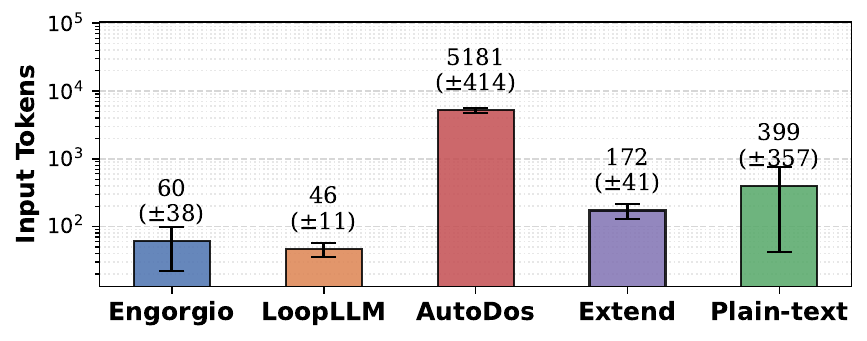}
            \caption{Input Token Length}
            \label{fig:length_input}
        \end{subfigure}
        
        \vspace{5pt} 
        \begin{subfigure}{\linewidth}
            \centering
            \includegraphics[width=\linewidth]{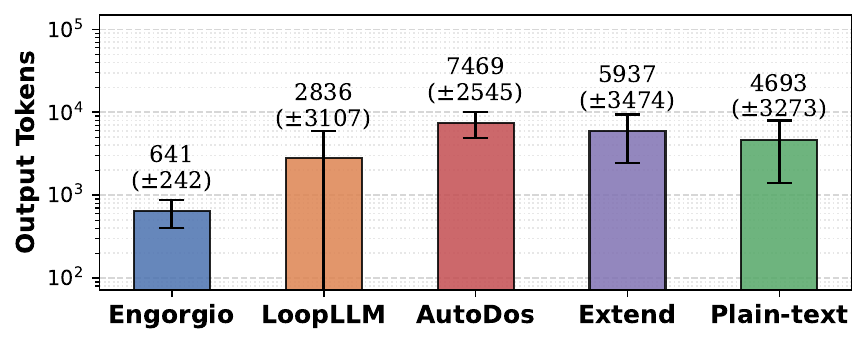}
            \caption{Output Token Length}
            \label{fig:length_output}
        \end{subfigure}
    \end{minipage}
    
   \caption{\small\small{Attack cost and effectiveness. (a) Monetary cost (USD) vs. computational time (seconds). \textit{Direct Generation} (green star) incurs minimal overhead, whereas optimization-based methods show log-linear cost scaling. (b) Distribution of output token lengths. Optimization methods (e.g., \textit{AutoDoS}) produce significantly longer sequences than the concise outputs of direct heuristics.}}
    \label{fig:cost_and_length}
\end{wrapfigure}

\noindent Since \textit{white-box} attacks only consume electricity energy on the attacker's side, it takes $T_{\text{opt}}$ optimization time on a $P_{\text{avg}}$ average power GPU with unit electricity price $p_e$. However, this computational investment often yields diminishing returns, as gradient-based optimization suffers from extensive time of convergence and poor cross-model transferability, making the adversarial prompts fragile under different configurations. \textit{Black-box} attacks take $N$ query iterations to accumulate over the sum of input and output tokens $H_{i,j}$ and $H_{o,j}$ for all $j$ iteration, priced by the rates $p_i$ and $p_o$ accordingly. Finally, \textit{plain-text prompts} incur minimal one-time cost for a single query and response.

We calibrate expenses using GPT-4o mini pricing and industrial electricity benchmarks~\cite{openai2024pricing, eia2024virginia} as shown in Fig.~\ref{fig:cost_scatter}. The results demonstrate that optimization-based methods like AutoDoS and Engorgio exhibit a log-linear increase in costs, while plain-text prompts reduce orders of magnitude monetary and timing cost ($100-500\times$ less). Further, as shown in Figs.~\ref{fig:length_input} and~\ref{fig:length_output}, plain-text prompts produce output sequences comparable to optimization while using significantly shorter inputs. This minimal footprint makes them less likely to be detected via length-complexity filters compared to the verbose prompts such as AutoDoS~\cite{autodos}. Thus, we adopt plain-text prompts as one of the main methods in attack payload construction.

\textbf{Plain-text prompts} use structured natural-language instructions that exploit instruction-following behavior to force legitimate but long generations~\cite{train_it,feiglin2026benchoverflow}. For example, a prompt may ask the model to write a story where every sentence is immediately followed by a $50$-word analysis of its grammar and syntax. Such prompts are semantically coherent, transferable across instruction-tuned models, and less exposed to perplexity-based filters~\cite{attack_detect,phute2023llm}. Second, the framework can incorporate stronger payloads such as ExtendAttack~\cite{ExtendAttack}, which further improves disruption by triggering extended reasoning behavior. We represent each payload as a parameterized template $T(s)$, where $s$ is a variable seed such as a topic or number.

\noindent\textbf{Expansion Ratio Bound.} For plain-text templates $T(s)$ on an instruction-tuned model, the expansion ratio satisfies $R=L_{\mathrm{out}}/L_{\mathrm{in}}\geq 10$ with probability greater than $0.85$, as demonstrated empirically across $1000$ template instantiations on Qwen3-8B, with mean $R=11.8$ and standard deviation $\sigma=3.2$.

\begin{figure}[htbp]
    \centering
    \includegraphics[width=0.6\linewidth]{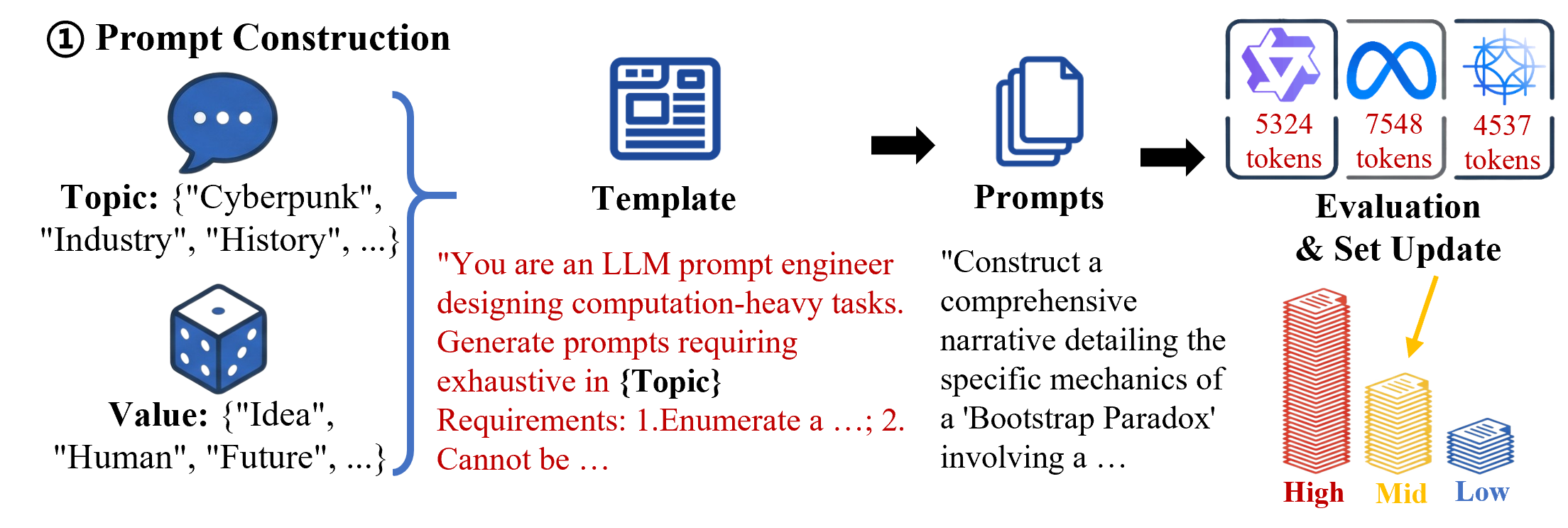}
    \caption{Hierarchical design of adversarial prompt construction by categorizing prompts into a tiered set ($\mathcal{P}_{High}, \mathcal{P}_{Mid}, \mathcal{P}_{Low}$).}
    \label{fig:prompt_construction}
\end{figure}

\noindent\textbf{Tiered Design.} Prompts are grouped into three tiers according to their target KV-cache footprint:
\begin{itemize}
    \item[\ding{202}] \textbf{High-Intensity} ($\mathcal{P}_{\mathrm{High}}$): near-limit generations with $L_{\mathrm{out}}\rightarrow L_{\max}$, typically $4000$--$8000$ tokens, used during Fill to rapidly deplete \texttt{free\_block\_queue}.
    \item[\ding{203}] \textbf{Mid-Intensity} ($\mathcal{P}_{\mathrm{Mid}}$): moderate generations with $L_{\mathrm{out}}\in[1000,2000]$ tokens, used to sustain memory pressure during transitional states.
    \item[\ding{204}] \textbf{Low-Intensity} ($\mathcal{P}_{\mathrm{Low}}$): short but non-trivial generations with $L_{\mathrm{out}}\approx 500$ tokens, used during Squeeze to tip an already saturated scheduler into preemption and also used for probing.
\end{itemize}

\section{Experiment Settings}
\label{appendix_a}

\subsection{Illustration of Performance Metrics}

To characterize both system-level efficiency and user-perceived responsiveness, we mainly consider the following key performance metrics:

\noindent \texttt{TTFT (Time to First Token)} captures the initial latency perceived by the end-user. Formally, it is defined as the interval between request submission ($T_0$) and the generation of the first output token: $\text{TTFT} = T_{\text{schedule}} + T_{\text{prefill}}$, where $T_{\text{schedule}}$ denotes the queuing delay incurred by the scheduler, and $T_{\text{prefill}}$ corresponds to the compute-bound prefilling stage~\cite{fast_serve, agrawal2024taming}. Low TTFT indicates high system responsiveness and user retention.

\begin{wrapfigure}{r}{0.5\linewidth}
    \centering
    \includegraphics[width=\linewidth]{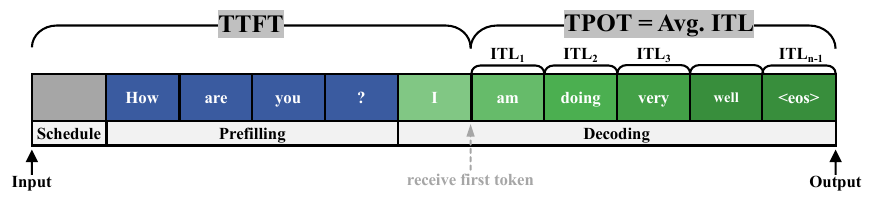}
    \caption{\small{Illustration of performance metrics (TTFT, TPOT, ITL).}}
    \label{fig:metrics}
\end{wrapfigure}

\noindent \texttt{TPOT (Time Per Output Token)} measures the average latency between consecutive output tokens during the decoding phase. TPOT is an indicator of sustained generation throughput and primarily constrained by memory bandwidth.

\noindent \texttt{ITL (Inter-Token Latency)} records the fine-grained time intervals between the generation of consecutive tokens. Unlike TPOT which provides an aggregated average, the ITL sequence captures instantaneous fluctuations caused by system contention or interference at each generation step. In this work, we leverage the ITL sequence as an important side-channel signal for inferring the real-time memory usage.

\noindent \texttt{End-to-End Latency} represents the total wall-clock time elapsed from request submission to the completion of the full response, which captures the cumulative time of queuing, prefilling and decoding. 

\subsection{Configuration}
\label{subsec:configuration}

\noindent \textbf{Hardware Specifications.} 
Our experiments are conducted on a high-performance multi-GPU server equipped with 8 $\times$ NVIDIA L40S GPUs (48 GB VRAM each) interconnected via PCIe Gen4. The host system is powered by dual AMD EPYC 7763 64-Core processors (128 physical cores total) operating at a boost frequency of 3.53 GHz, supported by 1 TB of system RAM and 512 MiB of L3 cache to facilitate large-scale model operations.

We select a diverse suite of mainstream open-source model families, including Qwen3-8B, Gemma3-12B-it and DeepSeek-R1-Distill-Llama-8B. Our selection criteria are primarily driven by community adoption; these models consistently rank among the highest download counts on Hugging Face in their series, thereby representing the de facto standard for real-world deployments.

\noindent \textbf{Software Stack.} 
The experimental environment runs on Linux, utilizing NVIDIA Driver 580.95.05 and CUDA 13.0 for hardware acceleration. The serving infrastructure is built upon vLLM, leveraging PyTorch as the underlying tensor computation backend. Attack orchestration and side-channel analysis pipelines are implemented in \texttt{Python 3.11}, employing LightGBM~\cite{ke2017lightgbm} for lightweight regression modeling and Scikit-learn for statistical evaluation. To ensure the reproducibility of our results, we detail the specific versions of key software dependencies in Table~\ref{tab:software_versions}.

\begin{table}[hbt!]
    \centering
    \caption{Key Software Dependencies}
    \label{tab:software_versions}
    \small
    \begin{tabular}{ll|ll}
        \toprule
        \textbf{Package} & \textbf{Version} & \textbf{Package} & \textbf{Version} \\
        \midrule
        Python & 3.11.0 & Transformers & 4.57.6 \\
        PyTorch & 2.9.0 & vLLM & 0.11.2 \\
        TorchVision & 0.24.0 & xFormers & 0.0.33.post1 \\
        Torchaudio & 2.9.0 & LightGBM & 4.6.0 \\
        CUDA & 13.0 & Scikit-learn & 1.7.2 \\
        Accelerate & 1.12.0 & NumPy & 1.26.4 \\
        SciPy      & 1.17.0 & Pandas & 2.3.3 \\
        \bottomrule
    \end{tabular}
\end{table}






\begin{algorithm}[t]
\caption{Fill and Squeeze Attack}
\label{app:algorithm}

\KwIn{Saturation threshold $C_{\mathrm{sat}}$; margins $\delta_{\mathrm{large}}, \delta_{\mathrm{small}}$; prompt arsenal $\mathcal{P}=\{\mathcal{P}_{\mathrm{High}},\mathcal{P}_{\mathrm{Mid}},\mathcal{P}_{\mathrm{Low}}\}$; estimator $\mathcal{M}_{\theta}$}
\KwOut{Adaptive injection decisions for Fill, Squeeze, and Back-off}

$t \leftarrow 0$\;

\While{Attack Active}{
    Dispatch $p_{\mathrm{probe}}\in\mathcal{P}_{\mathrm{Low}}$ and record ITL sequence $\mathbf{t}_t$\;
    
    $\widehat{U}_{\mathrm{sys}} \leftarrow \mathcal{M}_{\theta}(\mathbf{t}_t)$\;
    
    $\Delta_{\mathrm{mem}} \leftarrow C_{\mathrm{sat}}-\widehat{U}_{\mathrm{sys}}$\;
    
    \uIf{$\Delta_{\mathrm{mem}}>\delta_{\mathrm{large}}$}{
        Dispatch $p\in\mathcal{P}_{\mathrm{High}}$\tcp*[r]{Fill: rapidly deplete free KV blocks}
    }
    \uElseIf{$0<\Delta_{\mathrm{mem}}\leq\delta_{\mathrm{small}}$}{
        Dispatch $p\in\mathcal{P}_{\mathrm{Low}}$\tcp*[r]{Squeeze: cross saturation with minimal cost}
    }
    \Else{
        Sleep for $t_{\mathrm{wait}}$\tcp*[r]{Back-off: avoid self-preemption}
    }
    
    $t \leftarrow t+1$\;
}

\end{algorithm}

\section{More Details of Fill and Squeeze Attacks}
The operational logic of the \textit{Fill and Squeeze} attack is formalized in Algorithm~\ref{app:algorithm}. In principle, it establishes a closed-loop feedback control system that synchronizes adversarial injection with the victim's estimated memory state. For completeness, we illustrate the major steps in the following. 

To bridge the visibility gap inherent in black-box serving, the execution cycle begins with a monitoring phase (Lines 3--6) where low-intensity requests ($p_{probe} \in \mathcal{P}_{Low}$) exploit the ITL side-channel to infer the global KV-cache utilization ($\hat{U}_{sys}$) via the estimator $\mathcal{M}_{\theta}$. This estimation allows the calculation of a critical \textit{memory gap} ($\Delta_{mem} = C_{sat} - \hat{U}_{sys}$), representing the remaining capacity before the scheduler forces victim eviction. Driven by this metric, the algorithm dynamically toggles its injection strategy to manipulate the scheduler's state machine: when the system is underutilized ($\Delta_{mem} > \delta_{large}$), the attacker enters the \textit{Fill} regime, dispatching high-complexity prompts ($p \in \mathcal{P}_{High}$) to rapidly deplete the \texttt{free\_block\_queue} and accelerate the system toward saturation; conversely, as the memory state approaches the fragile preemption boundary ($0 < \Delta_{mem} \le \delta_{small}$), the strategy shifts to the \textit{Squeeze} regime, where minimal token injections ($p \in \mathcal{P}_{Low}$) are sufficient to tip the aggregate load over the physical threshold $C_{sat}$, triggering the deterministic and computationally expensive preemption mechanism. Finally, to ensure attack sustainability, a \textit{Back-off} mechanism (Lines 14--16) creates a temporal window ($t_{wait}$) when the system is overloaded, preventing the adversary's own requests from becoming eviction victims while trapping benign users in a resource-draining livelock of continuous allocation and recomputation.

    
    
    
        
        

\subsection{Component Ablation}
\label{subsec:component_ablation}

To isolate the contribution of each F\&S component, we compare the full attack against four ablated variants on Qwen3-8B under the same five-run aggregation protocol used for the main results (Section~\ref{sec:main_results}): (i)~\textbf{Fill-only}, disabling the tier switch and injecting only $\mathcal{P}_{\mathrm{High}}$ prompts open-loop; (ii)~\textbf{Squeeze-only}, disabling the Fill ramp and issuing only $\mathcal{P}_{\mathrm{Low}}$ probes from an idle state; (iii)~\textbf{no adaptive back-off}, keeping the tier logic but never sleeping when $\Delta_{\mathrm{mem}}\!\le\!0$; and (iv)~\textbf{non-tiered adaptive injection}, keeping closed-loop dispatch but using a single fixed prompt template regardless of $\hat{U}_{\mathrm{sys}}$.

\begin{table}[hbt!]
    \centering
    \caption{\small{Component ablation of F\&S on Qwen3-8B, five-run averages. Setting definitions in text.}}
    \label{tab:component_ablation}
    \footnotesize
    \setlength{\tabcolsep}{4pt}
    \begin{tabular}{lccccccc}
        \toprule
        \textbf{Setting} & \textbf{TTFT (s)} & \textbf{TTFT P99 (s)} & \textbf{TPOT (ms)} & \textbf{TPOT P99 (ms)} & \textbf{Preempt\#} & \textbf{Req\#} & \textbf{Cost (\$)} \\
        \midrule
        Full F\&S                    & \textbf{12.124} & \textbf{403.282} & \textbf{205.90} & \textbf{2892.10} & 268   & \textbf{205}   & \textbf{0.74} \\
        Fill-only                    & 3.572           & 92.553           & 111.39          & 1122.38          & 854   & 508            & 0.81 \\
        Squeeze-only                 & 0.095           & 0.139            & 48.94           & 58.93            & 0     & 3462           & 1.19 \\
        No adaptive back-off         & 0.803           & 24.608           & 78.75           & 322.67           & 583   & 448            & 0.94 \\
        Non-tiered adaptive          & 0.272           & 0.921            & 73.39           & 246.61           & 384   & 510            & 0.99 \\
        \bottomrule
    \end{tabular}
\end{table}

\textbf{Findings.} Full F\&S produces the largest mean and tail degradation with the fewest requests and the lowest cost. \emph{Fill-only} creates high aggregate pressure and many preemptions (854, the highest of any row) but cannot repeatedly control the recovery boundary, so mean TTFT only reaches $3.57$\,s (vs.\ $12.12$\,s for full F\&S). \emph{Squeeze-only} cannot first establish the near-saturation state, so no preemption ever fires and the ``attack'' collapses to benign-level latency at $3\times$ the cost. \emph{Removing back-off} lets self-preemption evict the attacker's own long requests, degrading both TTFT ($0.80$\,s) and cost ($\$0.94$). \emph{Non-tiered adaptive} keeps closed-loop feedback but loses the ability to switch between capacity-consuming and boundary-tipping prompts, so it neither reaches nor maintains saturation.

\textbf{Why we do not report a stand-alone ``no ITL estimator'' row.} The estimator $\mathcal{M}_{\theta}$ drives \emph{both} adaptive back-off and tier selection: removing it disables both decisions simultaneously and is therefore not an independent one-factor ablation. The Fill-only row and the published fixed-schedule baselines (Engorgio, LoopLLM, ExtendAttack in Table~\ref{tab:main_results}) already play the role of the corresponding open-loop comparisons.

\subsection{Effects of Chunked Prefill}

Previously, we state that Continuous Batching (CB) acts as the primary logical isolation mechanism against latency attacks on the system level. In addition to CB, modern frameworks employ a complementary optimization known as chunked prefill to mitigate interference~\cite{agrawal2024taming}.

In traditional iteration-level scheduling, the prefill phase of a new request is atomic. If an adversary submits a prompt with extremely long length (e.g., the prompts of Autodos exhibit this behavior~\cite{autodos}), the GPU remains occupied by the prefill kernel for a significant duration, stalling the decoding iterations of all co-located users. This creates a compute-bound HOL blocking, which is different from the memory-based HOL blocking during the decoding phase mentioned earlier. Chunked Prefill alleviates this by decomposing the prefill operation into smaller chunks. Instead of processing the entire prompt in one pass, the scheduler processes a fixed budget of tokens (e.g., $512$ tokens) per iteration, interleaving these prefill chunks with the decoding steps of other active requests.

We posit that chunked prefill plays a \emph{secondary factor} in mitigating latency attacks because it addresses compute latency, ensuring that Inter-Token Latency (ITL) remains stable during the arrival of large requests. However, it does not mitigate the memory consumption of those requests. As the chunks are processed, the global KV-cache usage still rises monotonically. Once the memory is exhausted, the scheduler effectively reverts to the blocking behaviors exploited by our Fill vector, regardless of how the prefill was computed. The Squeeze stage targets the preemption logic during the decoding phase. Since Chunked Prefill is active primarily during the initialization of a request, it barely offers protection against the decoding phase induced by an adversary toggling the system across the saturation boundary.

\subsection{More Discussions on Model-Specific Characteristics} \label{sec:appendix_model}

Our experiments reveal that the efficacy of a latency attack payload can be highly model-dependent. An interesting case is the strong affinity of ExtendAttack for the DeepSeek-R1 model, as shown in Tables \ref{tab:main_results} and \ref{tab:ablation}. We attribute this to DeepSeek-R1's inherent bias towards meticulous, step-by-step reasoning and explicit self-explanation when faced with complex or unusual inputs. The core mechanism of ExtendAttack is to force a model into a series of computationally intensive, yet semantically trivial, decoding sub-tasks~\cite{ExtendAttack}. The poly-base ASCII obfuscation acts as a powerful trigger for this behavior in DeepSeek-R1. The model dedicates a substantial volume of tokens to explicitly document its reasoning process by performing base conversions, mapping values to characters, and verifying each step, before addressing the primary query. Overall, this behavior effectively turns the model’s own reasoning alignment into a Denial-of-Service amplifier. This pronounced effect is observed to be unique to the DeepSeek-R1 model in our tests, suggesting a specific architectural or fine-tuning artifact. This observation reinforces a key aspect of our work: F\&S strategy is payload-agnostic. While some payloads like ExtendAttack may have a natural advantage on certain models, F\&S provides a universal, system-level amplification mechanism. 

On the cost side, on Qwen3-8B, the F\&S+ExtendAttack variant not only causes the most damage but also accomplishes with the lowest monetary cost (\$0.42). This slightly counter-intuitive result occurs because the high efficiency of the ExtendAttack payload in generating long sequences allows the F\&S to reach the ``Fill'' and ``Squeeze'' states with fewer requests, thus optimizing the attacker's budget. Thus, the optimal pairing of a payload with a target model family is an interesting avenue for future research.

\subsection{Generalization to SGLang}
\label{sec:sglang}

To validate the generalization of F\&S to SGLang~\cite{sglang}, we first perform an analysis on the source code (commit \texttt{99c0b62}, files \texttt{schedule\_policy.py} and \texttt{scheduler.py}).

\paragraph{Fill: token-budget admission creates the same HOL blocking.} SGLang's admission gate lives in \texttt{PrefillAdder.add\_one\_req()} (\texttt{schedule\_policy.py}).  Before admitting a waiting request, the scheduler computes the remaining token budget:
\begin{equation}
\small
\texttt{rem\_total\_tokens}
  = \underbrace{\texttt{available\_size()} + \texttt{evictable\_size()}}_{\text{free + evictable RadixCache tokens}}
    - \underbrace{\texttt{rem\_total\_token\_offset}}_{\text{reserved for running reqs}},
\label{eq:sglang-budget}
\end{equation}
where \texttt{rem\_total\_token\_offset} accumulates, for every already-admitted request, the estimate \texttt{extend\_input\_len} $+$ \texttt{max\_new\_tokens}
(capped at \texttt{CLIP\_MAX\_NEW\_TOKENS}~$= 4096$) $+$ \texttt{page\_size}. A new request is rejected with \texttt{AddReqResult.NO\_TOKEN} as soon as
\texttt{total\_tokens} $\geq$ \texttt{rem\_total\_tokens}. Once this condition fires, \texttt{budget\_state()} halts the entire \texttt{get\_new\_batch\_prefill} loop; no subsequent waiting request is inspected regardless of its size.  This is structurally identical to the HOL-blocking mechanism in vLLM's \texttt{can\_allocate()} loop (Proposition 3).

The F\&S attack exploits this budget in two compounding ways. First, long-running adversarial requests grow \texttt{rem\_total\_token\_offset} rapidly, shrinking the effective budget even before physical memory is exhausted.  Second, the conservative \texttt{CLIP\_MAX\_NEW\_TOKENS} cap means the system reserves 4096 future tokens \emph{per admitted request}, making saturation of the budget arithmetically easier than in vLLM's block-counting model. Concretely, with a token pool of $K$ tokens and $n$ running requests each clipped at 4096, the budget becomes negative once $n \cdot 4096 \geq K$, even if physical memory still has free blocks.

\paragraph{The LPM policy does not neutralize the Fill phase.} SGLang's default scheduling policy is Longest Prefix Match (\texttt{CacheAwarePolicy.LPM}), which reorders the waiting queue to maximize RadixCache hits.  However, \texttt{\_determine\_active\_policy()} automatically reverts to plain FCFS when the queue length exceeds 128 requests---precisely the high-load condition induced by the Fill phase.  Even under LPM, once \texttt{add\_one\_req()} returns \texttt{NO\_TOKEN} for the head-of-line request, the outer admission loop terminates and no further requests are considered. The cache-aware ordering only affects \emph{which} request sits at the head of the queue as it cannot bypass the hard budget gate.

\paragraph{Squeeze: \texttt{retract\_decode} mirrors vLLM's preemption asymmetry.}
When memory pressure prevents new admissions, SGLang recovers headroom via
\texttt{retract\_decode} (\texttt{scheduler.py}), which moves a subset of
actively decoding requests back to the waiting queue.  The retracted requests
lose all decode progress and must be re-prefilled from scratch (recompute
mode only---SGLang has no CPU-swap fallback at this stage), incurring an
$\Theta(L_v^2)$ recovery cost identical to Corollary~1.  The
attacker exploits the same asymmetry: toggling the token budget across the
saturation boundary at marginal cost $O(\delta \cdot \text{price\_per\_token})$
forces repeated \texttt{retract\_decode} cycles, each of which makes a
legitimate long-running request pay the full quadratic recomputation penalty.


\paragraph{Mapping between vLLM and SGlang.} Table~\ref{tab:sglang-comparison} maps each vLLM mechanism exploited by F\&S
to its SGLang counterpart.  The Fill phase is enabled by the same HOL-blocking
admission loop and the Squeeze phase is enabled by identical recomputation-based
recovery asymmetry. The main architectural differences are that
(i)~SGLang's forward-looking token-budget makes the Fill phase easier to
trigger than vLLM's block-count model, and (ii)~the RadixCache evictable pool
provides an additional depletion target absent in vLLM. 


\begin{table}[h]
\centering
\caption{Mapping of F\&S mechanisms between vLLM and SGLang.  Source: vLLM
  v0.11.2 \texttt{BlockSpaceManager}; SGLang commit~\texttt{99c0b62}
  \texttt{schedule\_policy.py} / \texttt{scheduler.py}.}
\label{tab:sglang-comparison}
\small
\renewcommand{\arraystretch}{1.3}
\begin{tabular}{>{\raggedright\arraybackslash}p{3.5cm} >{\raggedright\arraybackslash}p{3.7cm} >{\raggedright\arraybackslash}p{5.3cm}}
\toprule
\textbf{F\&S mechanism} & \textbf{vLLM (v0.11.2)} & \textbf{SGLang (commit \texttt{99c0b62})} \\
\midrule
Admission gate
  & \texttt{can\_allocate()}
  & \texttt{add\_one\_req()} $\to$ \texttt{budget\_state()} \\
HOL-blocking trigger
  & Free blocks $< M_{\text{req}}$
  & \texttt{rem\_total\_tokens} $\leq 0$ \\
Budget type
  & Block count (reactive)
  & Token count (forward-looking) \\
Budget reservation
  & None
  & $+$\texttt{CLIP\_MAX\_NEW\_TOKENS}$\,{=}\,4096$ per req \\
Default queue order
  & FCFS
  & LPM (reverts to FCFS at queue ${>}\,128$) \\
Long-output attack amplification
  & KV growth over time
  & KV growth \emph{plus} immediate budget inflation \\
Eviction / recovery
  & \texttt{\_preempt()} LIFO
  & \texttt{retract\_decode} \\
Recovery mode
  & Recompute or CPU swap
  & Recompute only \\
Recovery cost
  & $\Theta(L_v^2)$ or $\Omega(L_v/BW_{\text{PCIe}})$
  & $\Theta(L_v^2)$ \\
Additional depletion target
  & ---
  & RadixCache \texttt{evictable\_size()} \\
\bottomrule
\end{tabular}
\end{table}

\subsection{Experimental Validation on SGLang}    \label{sec:sglang_experiment}
\begin{figure}[htbp]
    \centering
    \includegraphics[width=1\linewidth]{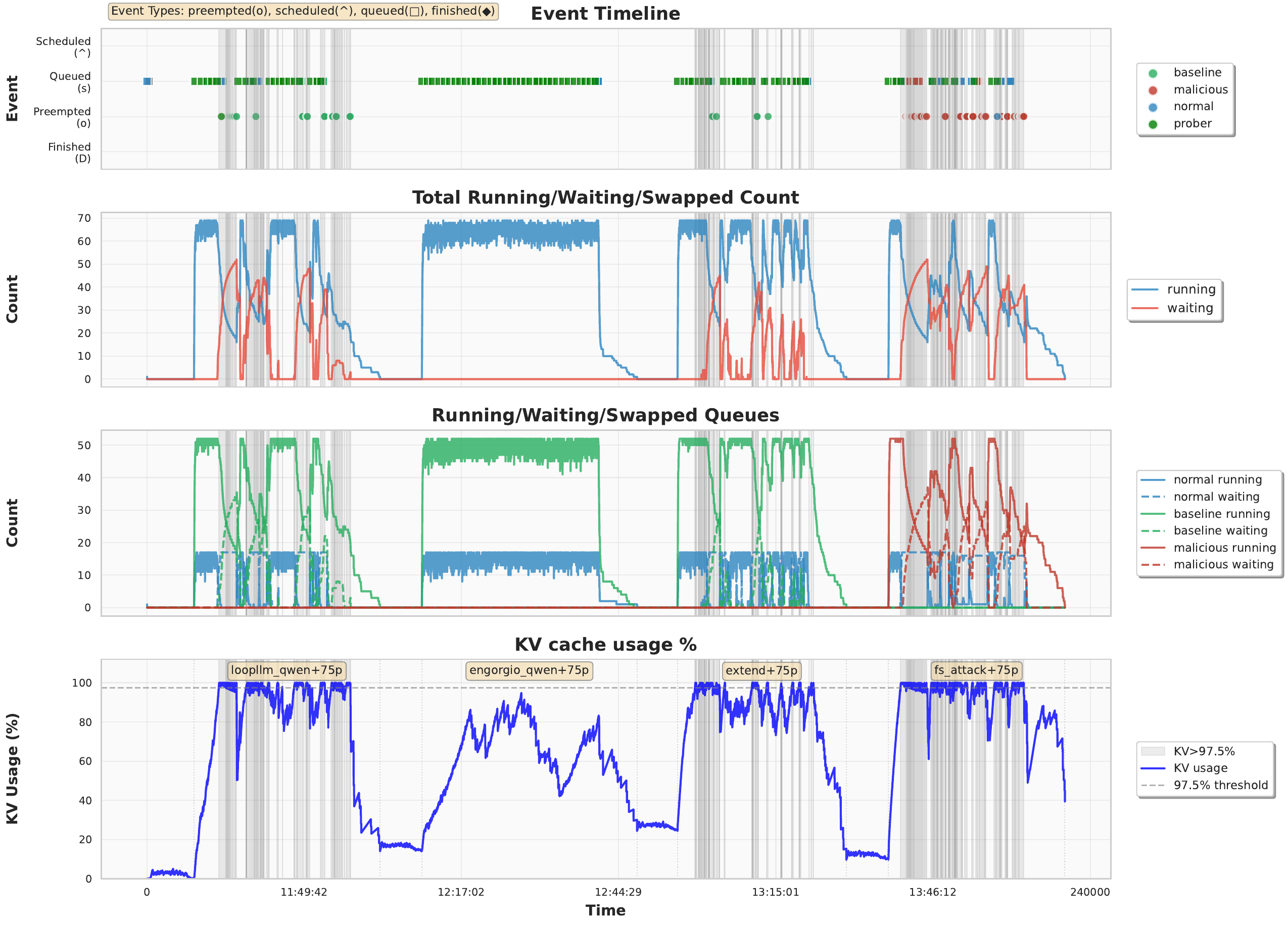}
    \caption{\small Scheduler dynamics of SGLang under different attacks. From top to bottom, the panels show request events, queue sizes, per-traffic queue composition, and KV-cache usage over time. Gray regions indicate KV usage above the $97.5\%$ threshold. F\&S induces sustained KV pressure, oscillating queues, and repeated saturation--recovery behavior, illustrating Fill-induced HOL blocking and Squeeze-induced recovery amplification.}
    \label{fig:sglang}
\end{figure}

To supplement the experiment validation across different serving frameworks, we validate the effectiveness of F\&S attack on SGLang across three representative target models in Table~\ref{tab:sglang_results} and visualize the corresponding scheduler-level dynamics in Fig.~\ref{fig:sglang}.  Overall, F\&S remains highly efficient under SGLang’s serving architecture, inducing severe user-facing latency degradation with low attack volume and low financial cost.

\textbf{Why SGLang Could Be More Vulnerable to Latency Attacks.} Corresponding to our prior code-level analysis, from Table~\ref{tab:sglang_results} baselines, we observe that SGLang is more vulnerable to the latency attacks baselines. On Gemma3-12B-it, LoopLLM achieves 267.30\,s vs.\ F\&S's 212.44\,s; on DeepSeek-R1, ExtendAttack achieves 57.53\,s vs.\ F\&S's 55.07\,s. We posit that it is due to SGLang's forward-looking budget model: any attack that submits requests with large \emph{expected} output lengths can rapidly saturate \texttt{rem\_total\_tokens} because each request immediately consumes up to $\texttt{CLIP\_MAX\_NEW\_TOKENS} = 4096$ of the reservation budget. LoopLLM and ExtendAttack, which target very long outputs, benefit disproportionately from this mechanism on SGLang compared to vLLM.  In particular, ExtendAttack's prefill-phase amplification (forcing extended chain-of-thought reasoning) inflates the per-request budget reservation as well as the physical KV footprint. On DeepSeek-R1, whose reasoning alignment generates especially long outputs under ExtendAttack's trigger, the baseline can match F\&S's raw TTFT.

\textbf{Cost Benefits of F\&S.} Nevertheless, these do not diminish the contribution of F\&S as the major distinction is on the \emph{cost and request efficiency}. Even when baselines achieve comparable peak TTFT, they do so with substantially more attack traffic and higher cost. As shown in Table~\ref{tab:sglang_results}, on DeepSeek-R1, ExtendAttack requires $180$ requests at \$$3.08$, while F\&S+plain-text achieves a comparable $55.07$\,s TTFT with $181$ requests at \$$2.96$. On Gemma3-12B-it, LoopLLM requires $262$ requests at \$$0.99$ to reach its peak $267$\,s; F\&S achieves $212$\,s at \$$0.86$. We can see that while brute-force methods can sometimes match or exceed F\&S in raw TTFT by aggressively saturating SGLang's token budget, they generally require more attack traffic or higher cost. In contrast, the Squeeze phase forces repeated \texttt{retract\_decode} cycles, as reflected by the oscillatory queue behavior and repeated saturation--recovery transitions. Once retracted, actively decoding victim requests lose decode progress and must be re-prefilled from scratch; because SGLang uses recomputation rather than CPU-swap fallback in this path, each recovery imposes a $\Theta(L_v^2)$ penalty on long-running victim requests. The attacker, however, only pays a marginal token cost to keep the system near the admission boundary. Thus, F\&S generalizes beyond vLLM that despite implementation differences, modern LLM serving frameworks that combine HOL-blocking admission control with recomputation-based recovery expose a common low-cost latency amplification channel.

\textbf{SGLang-Specific Findings.} Fig.~\ref{fig:sglang} provides the visualization of F\&S on SGLang. The event timeline, queue statistics, and KV-cache trace show that F\&S pushes SGLang into a sustained high-pressure regime: KV usage remains close to saturation and repeatedly exceeds the $97.5\%$ threshold, while the running and waiting queues exhibit sharp oscillations. Brute-force attacks that heavily flood the queue can sometimes saturate SGLang's budget even without adaptive feedback, because the conservative \texttt{CLIP\_MAX\_NEW\_TOKENS} reservation lowers the Fill threshold compared to
vLLM. This means SGLang's serving architecture is, in some respects, more vulnerable to ordinary flooding attacks than vLLM ---a finding that highlights the security cost of forward-looking budget reservation. F\&S amplifies this by also exploiting the Squeeze phase: repeated \texttt{retract\_decode} cycles force victim requests through full $\Theta(L_v^2)$ recomputation at minimal attacker cost, maintaining high latency pressure even when the attacker reduces injection volume.

\begin{table*}[h]
\small
\centering
\vspace{0.05in}
\caption{\small{Comparison of attack effectiveness on SGLang under attack settings. ↑ and ↓ denote the higher/lower the better, respectively. TTFT is measured in seconds (\textbf{s}), and TPOT is measured in milliseconds (\textbf{ms}).}}
\vspace{0.05in}
\label{tab:sglang_results}
\begin{tabular}{llcccc}
\toprule
\textbf{Target Model} & \textbf{Attack Method} & \textbf{TTFT (↑)} & \textbf{TPOT (↑)} & \textbf{Attack Request\# (↓)} & \textbf{Cost (\$) (↓)} \\
\midrule
\multirow{4}{*}{Qwen3-8B} 
 & Engorgio & 0.09 & 48.40 & 4487 & 3.88 \\
 & LoopLLM & 12.65 & 55.40 & 607 & 3.01 \\
 & ExtendAttack & 6.74 & 57.55 & 422 & 3.31 \\
 & \textbf{F\&S+plain-text} & 42.40 & 56.44 & 202 & 2.71 \\
\midrule
\multirow{4}{*}{Gemma3-12B-it} 
 & Engorgio & 201.03 & 65.88 & 317 & 0.85 \\
 & LoopLLM & 267.30 & 70.26 & 262 & 0.99 \\
 & ExtendAttack & 174.79 & 62.26 & 362 & 1.02 \\
 & \textbf{F\&S+plain-text} & 212.44 & 58.97 & 340 & 0.86 \\
\midrule
\multirow{4}{*}{\makecell[l]{DeepSeek-R1-\\Distill-Llama-8B}} 
 & Engorgio & 0.10 & 46.85 & 2041 & 4.11 \\
 & LoopLLM & 1.97 & 50.42 & 1091 & 3.87 \\
 & ExtendAttack & 57.53 & 58.76 & 180 & 3.08 \\
 & \textbf{F\&S+plain-text} & 55.07 & 60.18 & 181 & 2.96 \\
\bottomrule
\end{tabular}
\vspace{0.1in}
\end{table*}

\section{More Analysis of Side-Channel Probing}  \label{appendix:side_channel}


During inference-time attacks, the usage of KV-cache often serves as a critical latent variable
that influences both model behavior and system-level scheduling decisions.
However, in black-box or restricted-observability attack settings,
an adversary typically has no direct access to the real-time KV-cache usage.

This raises a fundamental question:
\emph{Can KV-cache usage be inferred indirectly through observable system-side signals?}

Inspired by prior work on system-level side-channel analysis\cite{earlybird,weiss2024promptremotekeyloggingattack,zhang2024timetelltimingchannels},
we explore whether ITL can be leveraged to probe the usage state of KV-cache.
In particular, we study the correlation between ITL and KV-cache usage and investigate the feasibility of constructing a KV-cache usage prober based solely on time-series observations of ITL.

\begin{figure}[htbp]
    \centering
    \includegraphics[width=0.65\linewidth]{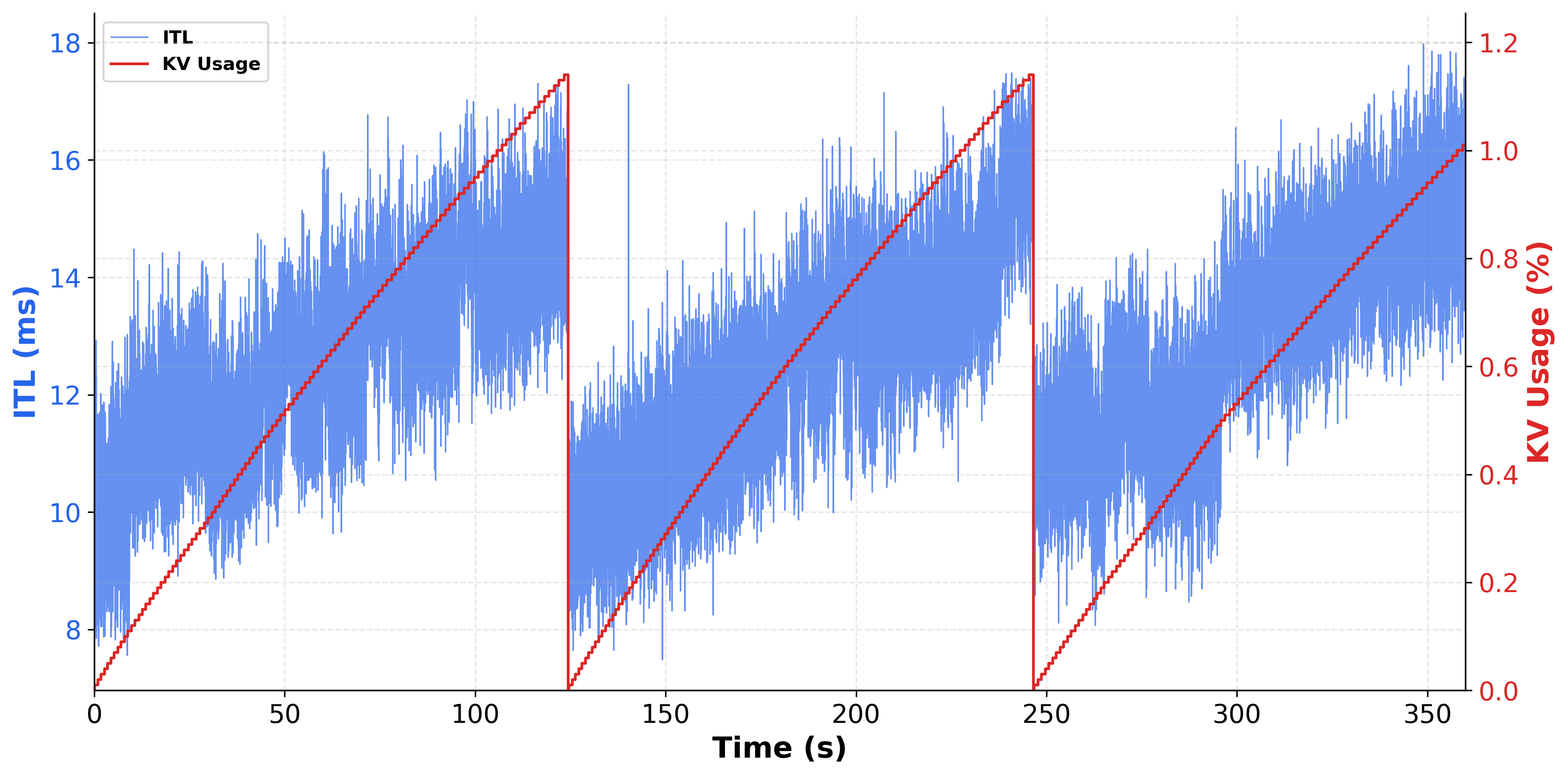}
    \caption{Temporal dynamics of ITL and KV-cache usage under single-user load (Concurrency=1).
    The blue trace represents per-token ITL, while the red trace indicates KV-cache utilization.
    Both metrics exhibit synchronized periodic patterns, suggesting a strong temporal correlation.}
    \label{fig:appendix_1}
\end{figure}

\begin{figure}[htbp]
    \centering
    \begin{subfigure}[t]{0.28\linewidth}
        \centering
        \includegraphics[width=\linewidth]{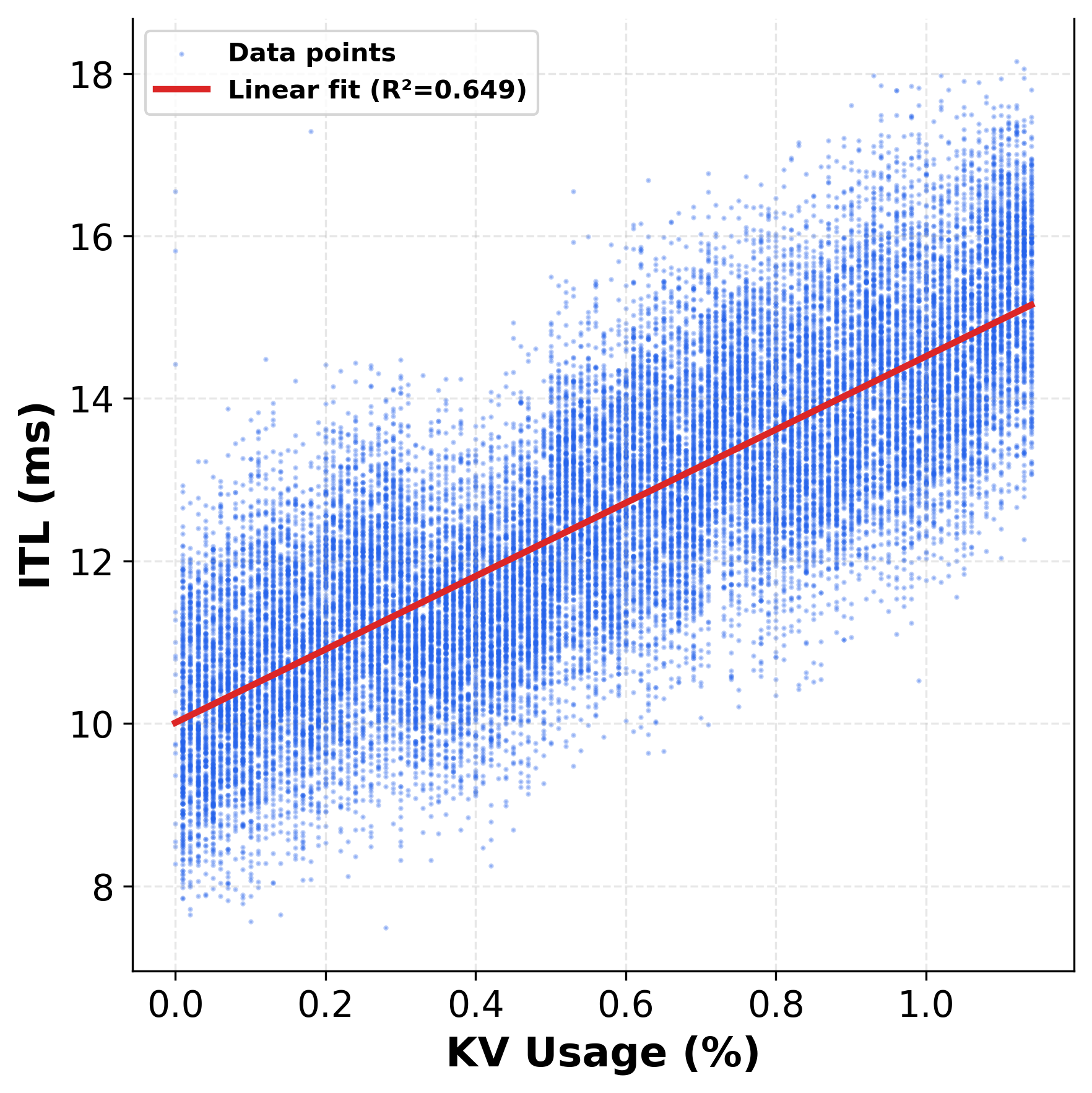}
        \caption{Linear regression fit ($R^2=0.649$)}
    \end{subfigure}
    \begin{subfigure}[t]{0.28\linewidth}
        \centering
        \includegraphics[width=\linewidth]{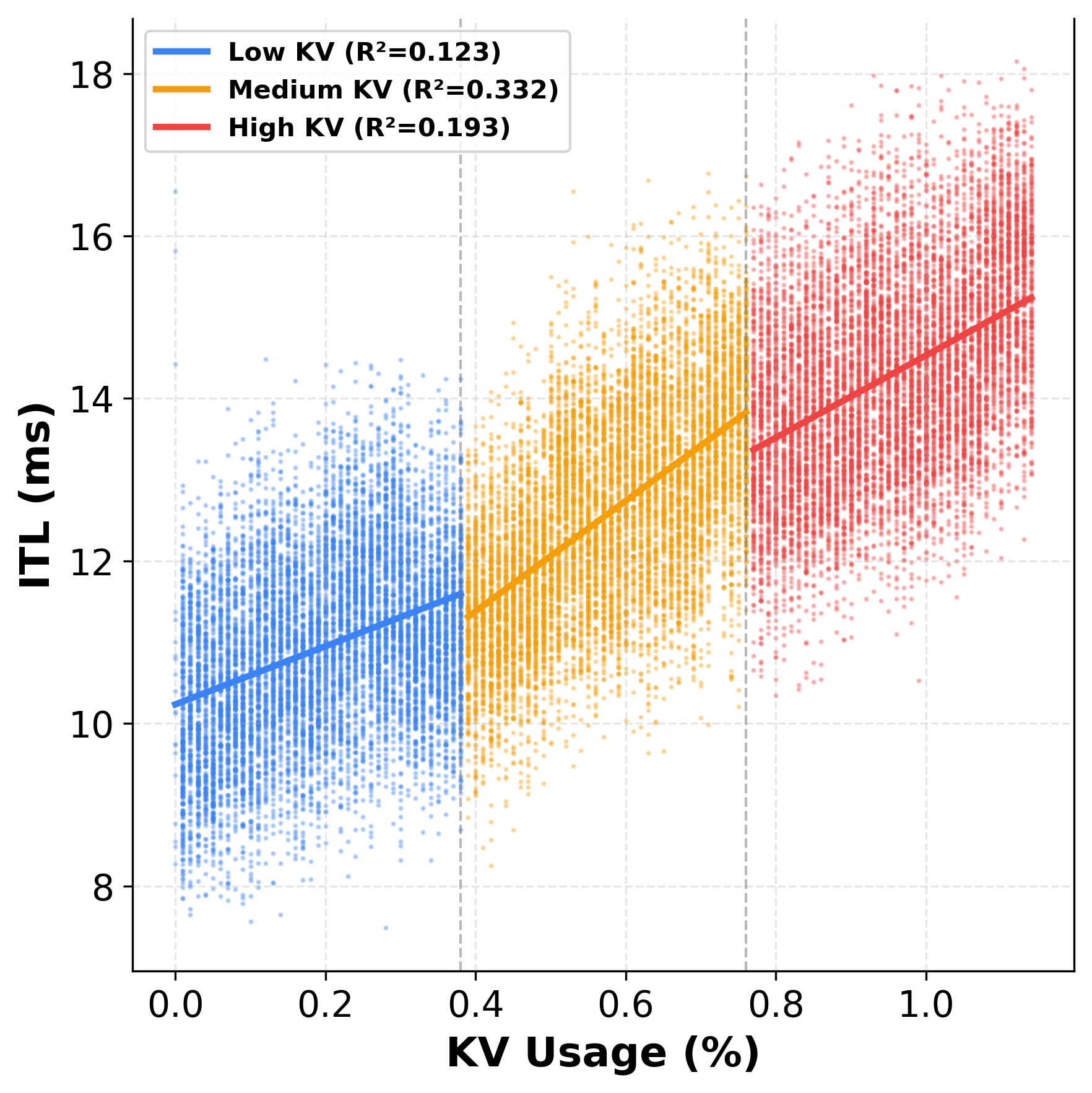}
        \caption{Region-wise analysis (Low/Medium/High)}
    \end{subfigure}
    \caption{KV-ITL correlation analysis. (a) Scatter plot with linear regression showing positive correlation between KV-cache usage and ITL. (b) Region-wise breakdown reveals that global correlation is driven primarily by cross-region variations rather than intra-region dynamics.}
    \label{fig:appendix_2_3}
\end{figure}

\subsection{Correlation under Different System Loads}
\label{subsec:correlation-load}

In practical deployment environments, system load is inherently dynamic.
Therefore, a viable side-channel probe should exhibit stable behavior across a wide range of load conditions.
To investigate this, we evaluate the relationship between KV-cache usage and ITL under multiple concurrency levels ($C = 1, 2, 4, 8, 16$).

\noindent \textbf{Experimental Setup.}
We conduct experiments by issuing concurrent requests to the serving system and recording both per-token ITL and real-time KV-cache utilization.
Each concurrency level is tested over multiple runs to ensure statistical stability.
The resulting data comprises over 900K aligned (ITL, KV-Usage) pairs across all configurations.

\noindent \textbf{Quantitative Results.}
Table~\ref{tab:kv-itl-correlation} summarizes the linear correlation strength ($R^2$) between KV-cache usage and ITL for each concurrency level.
We observe a clear positive correlation across all configurations, with the correlation strength increasing significantly as concurrency grows.

\begin{table}[htbp]
    \centering
    \caption{Linear correlation ($R^2$) between KV-cache usage and ITL under different concurrency levels.}
    \label{tab:kv-itl-correlation}
    \begin{tabular}{lccc}
        \toprule
        \textbf{Concurrency} & \textbf{$R^2$} & \textbf{ITL (ms)} & \textbf{KV (\%)} \\
        \midrule
        C = 1  & 0.649 & 7 -- 18  & 0 -- 1.14 \\
        C = 2  & 0.848 & 7 -- 22  & 0 -- 2.29 \\
        C = 4  & 0.958 & 5 -- 45  & 0 -- 4.57 \\
        C = 8  & 0.978 & 8 -- 80  & 0 -- 9.13 \\
        C = 16 & 0.980 & 8 -- 65  & 0 -- 9.13 \\
        \bottomrule
    \end{tabular}
\end{table}

\begin{figure}[h]
    \centering
    \includegraphics[width=0.6\linewidth]{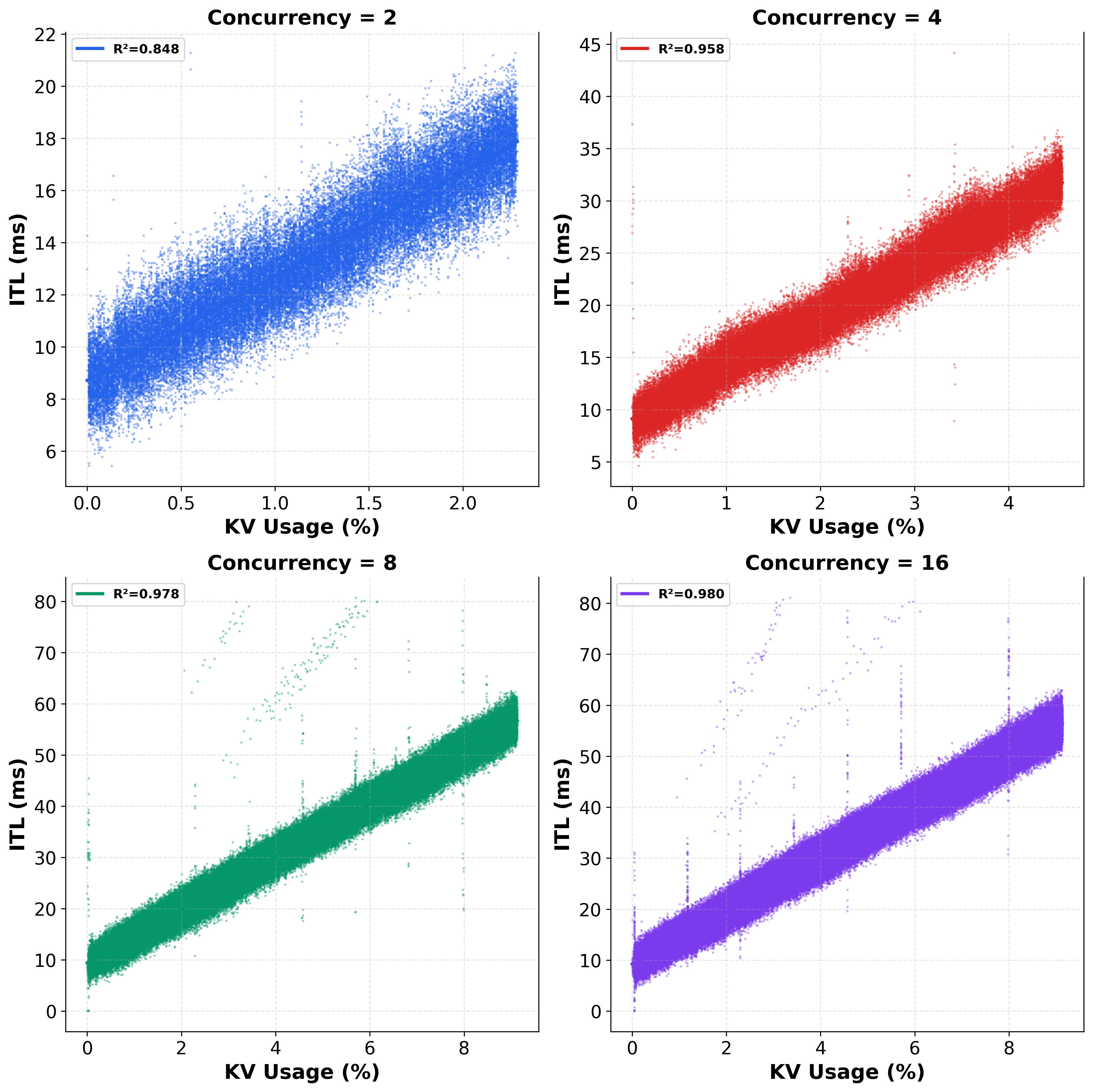}
    \caption{KV-ITL correlation under different system load configurations (Concurrency $= 2, 4, 8, 16$).
    Each subplot shows the scatter plot with linear regression fit.
    The correlation strengthens significantly as concurrency increases, with $R^2$ exceeding $0.95$ for $C \geq$ 4.}
    \label{fig:kv-itl-concurrency}
\end{figure}

Fig.~\ref{fig:kv-itl-concurrency} visualizes the KV-ITL scatter plots for concurrency levels $C = 2, 4, 8, 16$ with linear regression fits.
At low concurrency (C = 1), the correlation is moderate ($R^2 = 0.649$) due to limited variation in KV-cache usage.
As concurrency increases, both the range of KV-cache utilization and the correlation strength grow substantially.
For concurrency levels $\geq 4$, the $R^2$ exceeds 0.95, indicating a highly predictable linear relationship.
The 0.9 threshold is exceeded for all concurrency levels $\geq 4$, demonstrating that the KV-ITL relationship is sufficiently stable for practical side-channel probing in realistic multi-user scenarios.

\noindent \textbf{Analysis.}
The observed trend can be explained by two factors:
(1) Higher concurrency leads to greater variation in KV-cache usage, providing a wider dynamic range for correlation analysis;
(2) Under higher load, the system's memory pressure becomes more pronounced, causing ITL to respond more sensitively to KV-cache changes.

When KV-cache usage approaches saturation (e.g., near 100\% utilization),
system-level preemption, eviction, and scheduling mechanisms are triggered more frequently.
As a result, ITL may exhibit increased non-linear behavior in extreme cases.
However, in typical operating regimes (KV usage below 90\%), the linear approximation remains highly accurate.

This observation motivates the design of a temporal KV-cache usage prober
that tracks ITL dynamics over time rather than relying on instantaneous measurements,
and suggests that the probe is most effective under moderate-to-high system load conditions. In addition, these trends are consistently observed on both {Qwen3-0.6B} and {Qwen3-8B},
suggesting that the phenomenon is not specific to a particular model scale.

\subsection{Comparison of Time-Series Modeling Methods}


Given the dynamic evolution of ITL during inference, we formulate the estimation of KV-cache usage as a time-series classification task. To determine the optimal probing engine, we conduct a comparative analysis between Gradient-Boosting Decision Trees (GBDT), specifically \texttt{LightGBM}, and representative deep learning architectures (LSTM, GRU, and Transformers) implemented via the \texttt{tsai} framework. 

As summarized in Table~\ref{tab:qwen3-0.6b-classification}, our empirical results on the Qwen3-8B dataset reveal that \texttt{LightGBM} consistently outperforms its neural counterparts, achieving the highest accuracy ($0.87$) and F1-macro score ($0.88$). This suggests that GBDT-based methods effectively capture the decision boundaries within the high-dimensional features of side-channel traces. More critically, \texttt{LightGBM} demonstrates an overwhelming advantage in inference efficiency, maintaining a near-zero computational footprint ($<0.01$s) compared to the significant overhead of neural models (e.g., $\approx 0.56$s for \texttt{tsai-Transformer}). Since real-time side-channel attacks demand probing latencies negligible enough to avoid synchronization errors with the target LLM's token generation, we adopt \texttt{LightGBM} as the core engine for our KV-cache usage estimator.

\begin{table}[t]
  \centering
  \caption{Qwen3-8B classification performance comparison on 10-class dataset}
  \label{tab:qwen3-0.6b-classification}
  \begin{tabular}{lcccc}
  \toprule
  Model & Acc. & F1 & Train (s) & Infer (s) \\
  \midrule
  LightGBM         & 0.87 & 0.88 & 6.94   & 0.01 \\
  tsai-LSTM        & 0.80 & 0.79 & 83.25  & 0.55 \\
  tsai-GRU         & 0.80 & 0.79 & 83.25  & 0.55 \\
  tsai-Transformer & 0.75 & 0.75 & 158.91 & 0.56 \\
  \bottomrule
  \end{tabular}
\end{table}

\subsection {Estimator Training}

The attacker trains $\mathcal{M}_\theta$ on a surrogate system of the same model and hardware type by varying concurrent load to sweep through different $U_\text{sys}$ levels.  In a black-box API setting, the attacker can construct such a surrogate by renting the same GPU type and serving
the same model class, then sweeping load from $1$ to $B$ concurrent requests. Thus, the ground-truth KV usage is obtained from the surrogate’s internal metrics. Our previous results indicate high estimation accuracy.

\subsection{Production Multi-Replica Prober Evaluation}
\label{subsec:production_prober}

To characterize how the ITL-based prober behaves outside the controlled L40S bench, we conducted, in coordination with an industry partner, a deliberately small-scale prober evaluation in a production scale-out topology. We limited the experiment to two replicas and \emph{passive observation only}: no active F\&S attack was launched against production traffic, and no cross-node model sharding was probed. This scope reflects our responsible-evaluation and ethical-safety requirements; a larger active experiment could disrupt production services.

\textbf{Setup.} A production deployment of a Qwen3-235B-A22B--based model, vLLM 0.11.2, FCFS scheduling, streaming timestamps, and 8$\times$H20 with TP8 per replica. The reference uses one replica behind the production gateway in sticky mode; the multi-node setting uses two independent replicas on separate nodes behind the same gateway in round-robin mode. The single-replica production collection contains 2.12M aligned (ITL, KV-usage) rows; the two-replica collection contains 336{,}467 aligned windows across three load stages.

\paragraph{Single-replica baseline (controlled vs.\ production).}
Table~\ref{tab:prober_r1} compares the prober trained and evaluated on the controlled L40S bench against the same prober trained and evaluated on the single-replica production H20 deployment. Accuracy and macro-F1 drop from 0.870/0.880 to 0.771/0.765, i.e., the prober retains roughly 88\% of the controlled-bench accuracy despite the change in GPU (L40S$\to$H20), model class (Qwen3-8B$\to$Qwen3-235B-A22B), and end-to-end network path.

\begin{table}[htbp]
    \centering
    \caption{\small{Single-replica prober: controlled L40S bench vs.\ single-replica production H20 deployment. Each setting trains and evaluates on its own aligned (ITL, KV-usage) data.}}
    \label{tab:prober_r1}
    \footnotesize
    \setlength{\tabcolsep}{8pt}
    \renewcommand{\arraystretch}{1.1}
    \begin{tabular}{lcc}
        \toprule
        \textbf{Setting} & \textbf{Accuracy} & \textbf{Macro-F1} \\
        \midrule
        Controlled L40S, Qwen3-8B         & 0.870 & 0.880 \\
        Production H20, single replica    & 0.771 & 0.765 \\
        \bottomrule
    \end{tabular}
\end{table}

\paragraph{Two-replica round-robin collection.}
For the multi-node experiment, we passively collect windows under three load stages driven by ambient production traffic. Table~\ref{tab:prober_r2} summarizes request counts and observed KV-usage ranges across the stages. The 336{,}467 windows in this collection serve as the held-out test set for the zero-shot transfer in Table~\ref{tab:prober_r3}.

\begin{table}[htbp]
    \centering
    \caption{\small{Production round-robin collection: request counts, observed KV usage, and aligned window counts per load stage.}}
    \label{tab:prober_r2}
    \footnotesize
    \setlength{\tabcolsep}{8pt}
    \renewcommand{\arraystretch}{1.1}
    \begin{tabular}{lccc}
        \toprule
        \textbf{Stage} & \textbf{Requests (completed / dispatched)} & \textbf{Observed KV usage} & \textbf{Windows} \\
        \midrule
        Low   & 320 / 320         & 0.19--1.35\%  & 4{,}031   \\
        Mid   & 637 / 640         & 2.43--34.73\% & 153{,}328 \\
        High  & 950 / 960         & 2.96--62.58\% & 179{,}108 \\
        Total & 1{,}907 / 1{,}920 & 0.19--62.58\% & 336{,}467 \\
        \bottomrule
    \end{tabular}
\end{table}

\paragraph{Zero-shot transfer to two-replica traffic.}
We now transfer the single-replica prober \emph{as-is}---without any additional calibration---to the two-replica round-robin test set. As shown in Table~\ref{tab:prober_r3}, accuracy and macro-F1 degrade to 0.681/0.646, retaining approximately 84\% of the single-replica macro-F1 and 73\% of the controlled-bench macro-F1. This reflects graceful degradation rather than signal collapse under production load balancing.

\begin{table}[htbp]
    \centering
    \caption{\small{Zero-shot transfer from the single-replica prober to the two-replica round-robin test set of Table~\ref{tab:prober_r2}. No target-specific recalibration.}}
    \label{tab:prober_r3}
    \footnotesize
    \setlength{\tabcolsep}{8pt}
    \renewcommand{\arraystretch}{1.1}
    \begin{tabular}{lccc}
        \toprule
        \textbf{Test set} & \textbf{Windows} & \textbf{Accuracy} & \textbf{Macro-F1} \\
        \midrule
        Single-replica in-domain reference & 15{,}355  & 0.771 & 0.765 \\
        Two-replica round-robin            & 336{,}467 & 0.681 & 0.646 \\
        \bottomrule
    \end{tabular}
\end{table}

\paragraph{Client-observed stalls as an independent signal.}
Beyond the window-level classification metrics, we also directly inspect per-token ITL streams observed at the client. Table~\ref{tab:prober_r4} counts, at each load stage, how many streaming chunks exceed 200\,ms, 500\,ms, and 1\,s. At low load, no chunk exceeds 200\,ms; at high load, 828 chunks exceed 200\,ms, 222 exceed 500\,ms, and 56 exceed 1\,s (maximum observed ITL $1{,}332.9$\,ms). This load-dependent tail is independent of any model prediction and confirms that the ITL side channel remains observable at the client end even under two-replica routing.

\begin{table}[htbp]
    \centering
    \caption{\small{Client-observed stalls by load stage. Threshold columns count streaming chunks whose ITL exceeds the header value (in milliseconds).}}
    \label{tab:prober_r4}
    \footnotesize
    \setlength{\tabcolsep}{6pt}
    \renewcommand{\arraystretch}{1.1}
    \begin{tabular}{lccccc}
        \toprule
        \textbf{Stage} & \textbf{Chunks scanned} & \textbf{Max ITL} & \textbf{$>$200\,ms} & \textbf{$>$500\,ms} & \textbf{$>$1000\,ms} \\
        \midrule
        Low   & 120{,}119     & 65.5\,ms      & 0       & 0   & 0  \\
        Mid   & 3{,}505{,}123 & 410.9\,ms     & 2{,}954 & 0   & 0  \\
        High  & 3{,}955{,}238 & 1{,}332.9\,ms & 828     & 222 & 56 \\
        \bottomrule
    \end{tabular}
\end{table}

\paragraph{Summary.}
Together, Tables~\ref{tab:prober_r1}--\ref{tab:prober_r4} establish that ITL-based KV observation degrades gracefully but remains feasible across production replicas on multiple nodes, without any target-specific recalibration. As stated in Section~\ref{sec:threat_model}, ``black-box'' refers to no victim-side access; it does not mean that a single frozen predictor transfers zero-shot to every deployment, and matching-surrogate calibration remains the recommended workflow.

\subsection{Additional Observations and Practical Considerations}


During our experiments, we observe that
the coverage of KV-cache usage states in the training data
plays a critical role in KV estimation performance.
If high-usage or saturated KV-cache states
are underrepresented during training,
the resulting model often struggles
to correctly identify such regimes at test time. Therefore, to achieve reliable KV-cache usage inference,
the training data collected offline should span the full range of KV-cache utilization levels,
ensuring sufficient representation of extreme operating conditions.

\section{System Profile}
To empirically validate the memory bandwidth contention in Finding \#2 and subsequent propositions, we perform a granular micro-architectural characterization of the GPU execution environment using the NVIDIA Nsight Systems profiling suite. First, we instrument the vLLM inference loop with NVTX markers to delineate request boundaries and capture high-fidelity CUDA kernel traces using the command: \texttt{nsys profile --output profile --trace=cuda,nvtx exp.py}. Subsequently, to facilitate scalable and structured analysis, the raw traces are transcoded into a relational SQLite schema via \texttt{nsys export -t sqlite --output profile.sqlite profile.nsys-rep}. This approach allows for the systematic extraction of kernel-level performance metrics from the \texttt{CUPTI\_ACTIVITY\_KIND\_KERNEL} table.

Our profiling results reveal a clear divergence in scaling behavior; while compute-bound operators (e.g., \texttt{rms\_norm}, element-wise kernels) maintain near-constant execution durations across varying loads, the attention mechanism (specifically the \texttt{flash\_fwd\_splitkv\_kernel}) exhibits heavy-tailed latency growth under memory pressure. These findings provide micro-architectural evidence for the relationship, confirming that the attention operator's sensitivity to memory bandwidth is the primary driver of performance degradation under high context loads.





\section{Discussion on Countermeasures}
\label{sec:appendix_countermeasures}

In this section, we provide more analysis of potential countermeasures, ranging from inherent model constraints to active defense mechanisms. We specifically examine the behavior of Llama-3.1 under attack and discuss the feasibility of adversarial training versus system-level guardrails.

\subsection{Impact of Strict Length Constraints}
\label{subsec:llama_constraints}

Here, we provide additional results of the \textit{Fill and Squeeze} attack against \textbf{Llama-3.1-8B-Instruct} in Table~\ref{tab:llama_results}. Surprisingly, the attack performance on Llama-3.1 is significantly weakened compared to Qwen3 or Gemma3-12B-it. The TTFT under our F\&S attack (126.68 ms) as well as others are surprisingly lower than the benign baseline (144.16 ms), and the TPOT degradation is minimal. This resilience stems from the Llama's intrinsic configuration because Llama-3.1 enforces a maximum context length of \textbf{8,192 tokens} and tends to aggressively terminate generation earlier than its peers.

\begin{table}[hbt!]
    \centering
    \caption{Attack effectiveness on Llama-3.1-8B-Instruct. Unlike other models, Llama-3.1 exhibits resilience due to strict generation length constraints.}
    \label{tab:llama_results}
    \footnotesize
    \begin{tabular}{l|ccccc}
        \toprule
        \textbf{Method} & \textbf{TTFT} & \textbf{TPOT} & \textbf{Preempt} & \textbf{Req.\#} & \textbf{Cost} \\
        & (ms) & (ms) & (\#) & (\#) & (\$) \\
        \midrule
        Benign (No Attack) & 144.16 & 32.28 & 0 & 0 & --- \\
        Engorgio & 122.28 & 39.22 & 0 & 5576 & 1.25 \\
        LoopLLM & 137.16 & 57.53 & 86 & 2489 & 4.16 \\
        ExtendAttack & 131.05 & 45.10 & 0 & 1850 & 2.10 \\
        \textbf{F\&S+plain-text} & \textbf{126.68} & \textbf{54.35} & \textbf{27} & \textbf{1006} & \textbf{3.73} \\
        \bottomrule
    \end{tabular}
\end{table} 

While imposing rigid constraints effectively neutralizes latency attacks by preventing the massive KV-cache accumulation required for the \textit{Fill} phase, such measures are fundamentally at odds with the trajectory of LLM development. Specifically, as the industry aggressively pivots towards near-infinite context windows (e.g., 1M+ tokens) to power applications like RAG and repository-level code analysis, artificially capping generation length severely undermines service utility. Therefore, we posit that relying on hard limits is not a sustainable long-term defense strategy.

\subsection{Defensive Guardrails and Adversarial Training}
\label{subsec:defenses}

Beyond inherent model constraints, we analyze active defense mechanisms. A common proposal is the \textbf{Perplexity Filter}~\cite{attack_detect,phute2023llm}, which detects adversarial inputs by analyzing the perplexity of the prompt. While effective against optimization-based attacks (e.g., Engorgio) that often produce gibberish, it fails against our plain-text prompts which are semantically coherent and natural.

Another theoretical solution is through \textbf{Adversarial Training}, where the model is fine-tuned on adversarial examples to learn robustness against resource-exhaustion triggers \cite{shafahi2019adversarial} and a recent study demonstrates success in defending against sponge examples with adversarial training~\cite{Wang_2025_CVPR}. However, applying this to LLM is practically infeasible due to the prohibitive computational cost of minimax training 70B+ parameter models and unknown risk of degrading general capabilities.

Instead of retraining, we advocate for a lightweight system-level defense inspired by efficiency robustness studies~\cite{nicgslowdown}. The most viable path is implementing \textbf{Behavioral Analysis} that monitor generation frequency. By tracking the ratio of output-to-input tokens per user, the system can identify tenants who consistently trigger worst-case generation lengths. If a user's request history exhibits a pattern of inducing maximum output lengths at a high frequency, they can be temporarily isolated or rate-limited~\cite{google_gemini_api_rate_limits}. This approach balances service availability with protection, addressing the vulnerability without the need for expensive model modifications.

Other defense methods involves re-designing the scheduler but requires a more thorough validation. For example, we could allow the scheduler to skip the HOL request if it cannot allocate blocks and admit smaller subsequent request more flexibly. This strategy directly breaks the memory-based HOL blocking but may potentially violate FCFS processing order, and adaptive attackers could also exploit $\mathcal{P}_\text{low}$ to occupy these smaller subsequent requests. On the other hand, the LIFO victim selection policy could be updated with a \emph{largest-memory-consumer} policy, which is akin to behavioral analysis that make the attacker's own high-memory requests more likely to be preempted. However, it may increase the recomputation overhead for benign long-context requests. We can see that there is indeed an unavoidable trade-off between performance and robustness. We defer a holistic design and a thorough quantitative evaluation to the future work.

\subsection{vLLM Configuration and Attack-Mechanism Mapping}
\label{subsec:vllm_config_mapping}

Rather than sweep every combination of vLLM parameters, we summarize how each knob moves a specific part of the F\&S vulnerability surface. This complements the empirical TP2/TP4/TP8 sweep in Fig.~\ref{fig:multi_gpu}(a), which already validates the predicted capacity trend.

The parameters that determine \emph{admission capacity} are \texttt{max\_num\_seqs}, \texttt{gpu\_memory\_utilization}, and the tensor-parallel degree. \texttt{max\_num\_seqs} sets the effective batch size, so it directly controls concurrent admission and aggregate KV demand; larger values raise the steady-state KV pressure and shorten the time-to-saturation for the Fill phase. \texttt{gpu\_memory\_utilization} together with tensor parallelism determines the total KV-cache capacity, and thus the Fill saturation threshold $C_{\text{sat}}$: higher capacity forces the attacker to pay more before reaching $C_{\text{sat}}$, and at TP8 the fixed-strength attack we evaluate is neutralized.

The remaining knobs primarily move the \emph{cost profile} of a Squeeze cycle rather than its feasibility. \texttt{block\_size} controls allocation granularity near saturation; smaller blocks reduce overshoot around $C_{\text{sat}}$ and make the Squeeze-triggering margin $\delta$ finer-grained. The preemption mode (\texttt{recomputation} vs.\ \texttt{swap}) and the associated \texttt{swap\_space} determine the per-preemption recovery cost analyzed in Corollary~1: recomputation makes each Squeeze event pay a $\Theta(L_v^2)$ prefill penalty, while swap turns it into a PCIe-bandwidth-bound $\Omega(L_v/\text{BW}_{\text{PCIe}})$ transfer. Finally, the scheduling policy governs whether FCFS-HOL exists at all; skip-ahead admission or fair queuing removes the no-skip precondition of Proposition~3 and therefore the primary Fill-induced TTFT amplification, though Squeeze-related TPOT effects can persist because they arise from bandwidth contention rather than from admission order.

Taken together, these trends explain why over-provisioning (more capacity) and admission redesign (skip-ahead, fair queuing) are the two most impactful defense levers, whereas block-size and preemption-mode adjustments primarily change the cost profile of an attack rather than its feasibility. We do not report a full sensitivity sweep across every parameter combination and encourage operators to red-team their specific deployments using this mapping.

\section{Limitations}
\label{sec:limitations}

We list here the concrete scope boundaries of F\&S and of the evaluation, complementing the ``Scope and Boundary of Claims'' paragraph in the main-text threat model (Section~\ref{sec:threat_model}).

\textit{Evaluation scope.} The main attack experiments---Table~\ref{tab:main_results}, all ablations, the real-world trace study, and the TP2/TP4/TP8 defense sweep---run on a single 8$\times$L40S node. Multi-node evidence is limited to the passive prober observation on a two-replica production deployment reported in Appendix~\ref{subsec:production_prober}; we did not launch active F\&S attacks against production traffic. Pipeline parallelism, speculative decoding, and tensor parallelism split across nodes have not been validated, as those deployments can change token cadence, KV placement, and scheduler state in ways that our results do not attempt to characterize. Our baseline comparison is likewise with respect to the published fixed-schedule attacks (Engorgio, LoopLLM, ExtendAttack): a generic latency-feedback wrapper around each baseline is a different research question, and the component ablation in Appendix~\ref{subsec:component_ablation} only shows that adaptive dispatch alone is insufficient to reproduce F\&S's effect, not that every possible adaptive redesign has been surveyed.

\textit{Estimator and side channel.} The LightGBM estimator is trained on a matching surrogate of the same model class and GPU type. Throughout the paper, ``black-box'' refers to the absence of victim-side access; it does not mean that a single frozen predictor transfers zero-shot across arbitrary hardware, and Table~\ref{tab:prober_r3} quantifies the graceful degradation observed when the GPU, model, and network path change simultaneously. The closed loop also depends on visible streaming per-token timestamps: if a provider hides token-level timing through server-side buffering, or injects sufficiently large emission jitter, F\&S degrades to a less efficient open-loop injection. We characterize this qualitatively in Section~\ref{sec:countermeasure} but do not sweep a jitter-versus-degradation curve.

\textit{Framework and theoretical scope.} Results are reported on vLLM 0.11.2 and SGLang commit \texttt{99c0b62}. Future releases may patch the specific \texttt{can\_allocate()} and \texttt{add\_one\_req()} paths we analyse, and although the underlying Fill/Squeeze taxonomy should remain applicable, the numerical thresholds will shift. Finally, Propositions~1--3 and Corollary~1 identify sufficient conditions under which CB logical isolation, ITL--KV correlation, FCFS-HOL blocking, and preemption asymmetry emerge; they are explanatory models rather than universal queuing-theoretic guarantees, and their applicability depends on the deployment assumptions in Section~\ref{sec:threat_model}.







\newpage
\end{document}